\documentclass[aps,prb,a4paper,twocolumn,10pt,superscriptaddress,notitlepage]{revtex4-2}
\newcommand{\papertitle}{Cryogenic nano-imaging of second-order moiré superlattices}

\def \pathfigs {Figures}
\def \scalefig {0.97}
\def \scalefigfive {0.97}
\def \scalefigtransport {1}

\newcommand{\NT}[1]{{\color{black}#1}}

\newcommand{\NTblue}[1]{{\color{blue}#1}}

\usepackage[utf8]{inputenc}
\usepackage{graphicx}
\usepackage{graphicx}
\usepackage{amsmath}
\usepackage{amssymb}
\usepackage{natbib}
\usepackage{float}
\usepackage{tikz}
\usepackage{siunitx}
\usepackage{dcolumn}
\usepackage{lmodern}
\usepackage[T1]{fontenc}
\usepackage[unicode=true,
	    colorlinks=true,
	    linkcolor=blue,
	    citecolor=blue,
	    urlcolor=blue]{hyperref} 
\usepackage{sansmath,caption}
\DeclareCaptionLabelSeparator{vertline}{ \textnormal{\textbar} }
\usepackage[font=sf, labelfont={sf,bf},size=small,justification=RaggedRight,textfont={sf,sansmath},labelsep=vertline]{caption}
\usepackage{ulem}

\usepackage{xcolor}

\usepackage{chemformula}
\usepackage{empheq}
\usepackage{comment}
\usepackage{bm}
\usepackage{gensymb}

\usepackage[misc]{ifsym}
\usepackage{fontawesome}

\setcitestyle{super}

\usepackage{xr}
\makeatletter

\newcommand*{\addFileDependency}[1]{
\typeout{(#1)}
\@addtofilelist{#1}
\IfFileExists{#1}{}{\typeout{No file #1.}}
}\makeatother

\newcommand*{\myexternaldocument}[1]{%
\externaldocument{#1}%
\addFileDependency{#1.tex}%
\addFileDependency{#1.aux}%
}

\myexternaldocument{supplement}

\usepackage[top=2.5cm,bottom=2.5cm,left=1.5cm,right=1.5cm]{geometry}

\begin{document}

\title{\Large\textsf{\papertitle}}

\author{Niels C.H. Hesp}
\affiliation{\footnotesize ICFO-Institut de Ci\`{e}ncies Fot\`{o}niques, The Barcelona Institute of Science and Technology, Av. Carl Friedrich Gauss 3, 08860 Castelldefels (Barcelona),~Spain}

\author{Sergi Batlle-Porro}
\affiliation{\footnotesize ICFO-Institut de Ci\`{e}ncies Fot\`{o}niques, The Barcelona Institute of Science and Technology, Av. Carl Friedrich Gauss 3, 08860 Castelldefels (Barcelona),~Spain}

\author{Roshan Krishna Kumar}
\affiliation{\footnotesize ICFO-Institut de Ci\`{e}ncies Fot\`{o}niques, The Barcelona Institute of Science and Technology, Av. Carl Friedrich Gauss 3, 08860 Castelldefels (Barcelona),~Spain}

\author{Hitesh Agarwal}
\affiliation{\footnotesize ICFO-Institut de Ci\`{e}ncies Fot\`{o}niques, The Barcelona Institute of Science and Technology, Av. Carl Friedrich Gauss 3, 08860 Castelldefels (Barcelona),~Spain}

\author{David Barcons-Ruiz}
\affiliation{\footnotesize ICFO-Institut de Ci\`{e}ncies Fot\`{o}niques, The Barcelona Institute of Science and Technology, Av. Carl Friedrich Gauss 3, 08860 Castelldefels (Barcelona),~Spain}

\author{Hanan Herzig Sheinfux}
\affiliation{\footnotesize ICFO-Institut de Ci\`{e}ncies Fot\`{o}niques, The Barcelona Institute of Science and Technology, Av. Carl Friedrich Gauss 3, 08860 Castelldefels (Barcelona),~Spain}

\author{Kenji Watanabe}
\affiliation{\footnotesize Research Center for Functional Materials, National Institute for Materials Science, 1-1 Namiki, Tsukuba 305-0044,~Japan}

\author{Takashi Taniguchi}
\affiliation{\footnotesize International Center for Materials Nanoarchitectonics, National Institute for Materials Science,  1-1 Namiki, Tsukuba 305-0044,~Japan}

\author{Petr Stepanov}
\email{pstepano@nd.edu}
\affiliation{\footnotesize ICFO-Institut de Ci\`{e}ncies Fot\`{o}niques, The Barcelona Institute of Science and Technology, Av. Carl Friedrich Gauss 3, 08860 Castelldefels (Barcelona),~Spain}
\affiliation{\footnotesize Department of Physics and Astronomy, University of Notre Dame, Notre Dame, IN 46556,~USA}
\affiliation{\footnotesize Stavropoulos Center for Complex Quantum Matter, University of Notre Dame, Notre Dame, IN 46556,~USA}

\author{Frank H.L. Koppens}
\email{frank.koppens@icfo.eu}
\affiliation{\footnotesize ICFO-Institut de Ci\`{e}ncies Fot\`{o}niques, The Barcelona Institute of Science and Technology, Av. Carl Friedrich Gauss 3, 08860 Castelldefels (Barcelona),~Spain}
\affiliation{\footnotesize ICREA-Instituci\'{o} Catalana de Recerca i Estudis Avan\c{c}ats, 08010 Barcelona, Spain}

\maketitle

\noindent

\noindent
\textbf{Second-order superlattices form when moiré superlattices of similar periodicities interfere with each other, leading to even larger superlattice periodicities. These crystalline structures have been engineered utilizing two-dimensional (2D) materials such as graphene and hexagonal boron nitride (hBN) under specific alignment conditions. Such specific alignment has shown to play a crucial role in facilitating  correlation-driven topological phases featuring the quantized anomalous Hall effect. While signatures of second-order superlattices have been identified in magnetotransport experiments, any real-space visualization is lacking to date. In this work, we present \NT{electronic transport measurements and cryogenic nanoscale photovoltage (PV) measurements} that reveal a second-order superlattice in magic-angle twisted bilayer graphene closely aligned to hBN. This is evidenced by long-range periodic photovoltage modulations across the entire sample backed by the corresponding electronic transport features. Supported by theoretical calculations, our experimental data show that even minuscule strain and twist-angle variations on the order of 0.01$\degree$ can lead to a drastic change of the second-order superlattice structure between local one-dimensional, square or triangular types. Our real-space observations therefore serve as a strong `magnifying glass' for strain and twist angle and can shed new light on the mechanisms responsible for the breaking of spatial symmetries in twisted bilayer graphene, and pave an avenue to engineer long-range superlattice structures in 2D materials using strain fields.}

\begin{figure*}[t]
    \centering
    \includegraphics{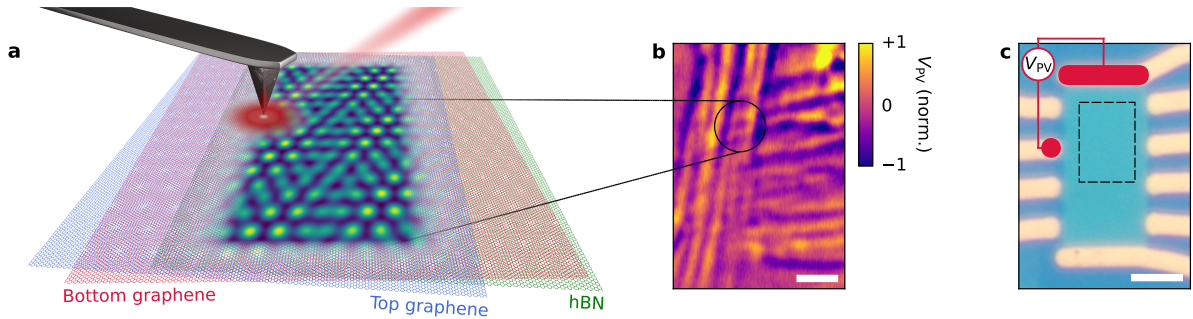}
    \caption{\textbf{Nanoscale photovoltage measurements of MATBG/hBN SOSLs at \textit{T}=10~K.} \textbf{a}, \NT{Schematic illustration of the formation of a SOSL and the experimental design. The hBN/bottom graphene and the two graphene layers form a first-order superlattice. The interference of these two short-range superlattices is captured by the long-range periodicity that encompasses a SOSL. In our experiment, a metal-coated AFM tip is positioned in the focus of an infrared laser beam, creating a hot-spot underneath the tip, which in turn locally heats the charge carriers in the MATBG.} Any modulation of the electronic properties can induce a global photovoltage that we read out with the device contacts. \textbf{b}, An example of a photovoltage measurement on our device in the black dashed area in \textbf{c}, revealing two sets of fringes that we associate with the SOSL potential. The excitation energy is 116~meV and the scale bar is 1~$\mu$m. \textbf{c}, An optical micrograph of the studied sample. The red lines indicate the probes used for the photovoltage measurements in \textbf{b}. The scale bar is 4~$\mu$m.}
    \label{fig:intro-experiment}  
\end{figure*}

\begin{figure*}[t]
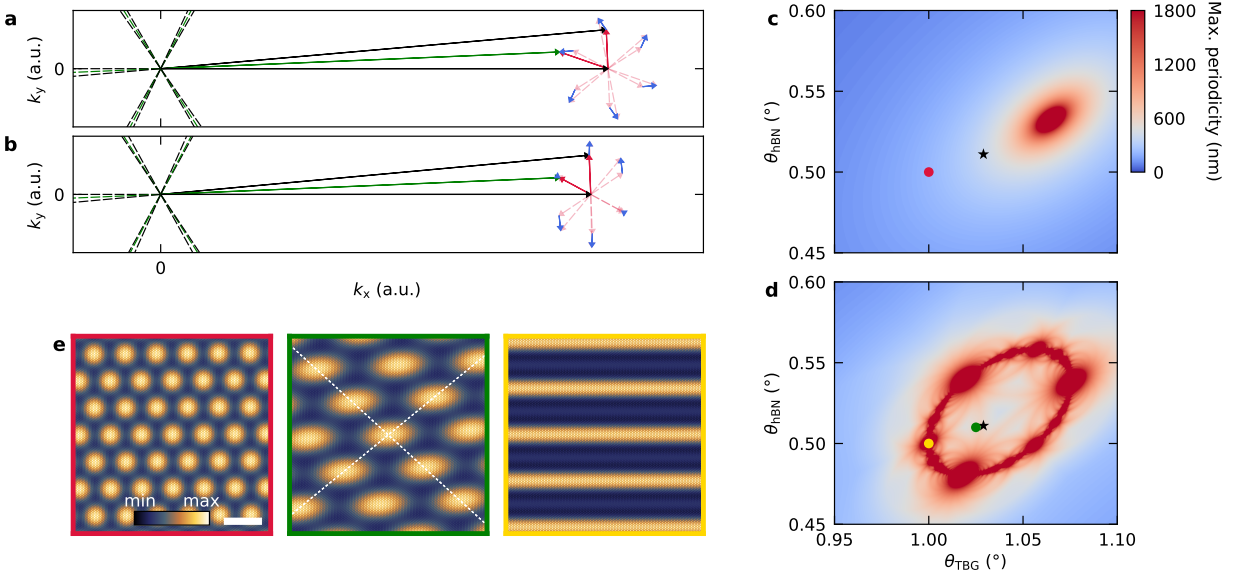

    \centering
    \raisebox{-\height}{\includegraphics[scale=\scalefig]{\pathfigs /Fig4abe.pdf}}
    \raisebox{-\height}{\includegraphics[scale=\scalefig]{\pathfigs /Fig4cd-second-order-periodicity-calculation.pdf}}

    \caption{\textbf{Calculation of second order superlattice properties.} \textbf{a}, Reciprocal lattice vectors of the unstrained and closely aligned hBN (green) and graphene layers (black). The corresponding dashed vectors account for lattice periodicities in other directions (at multiples of $60 \degree$). The red vectors denote the hBN-Gr and TBG superlattices, and their equivalents at multiples of $60 \degree$ (dashed red vectors). The resultant SOSL reciprocal vectors (blue) appear from the summation of red vectors. \textbf{b}, Same reciprocal vectors with uniaxial strain applied to the graphene lattices.  \textbf{c}, SOSL periodicity as a function of $\theta_{\rm TBG}$ and $\theta_{\rm hBN}$ of the unstrained SOSL lattice (corresponding to \textbf{a}). \textbf{d}, Maximum SOSL periodicity (along one of its principal axes) of the deformed lattice when strain applied to the graphene layers (corresponding to \textbf{b}). The star indicates a data point obtained from the electronic transport measurements. \textbf{e}, Three examples of different SOSL geometries: a triangular lattice, a square lattice, and a 1D lattice. The colors of the frames corresponds to the color of the dots in \textbf{d}). The color bar represents the calculated potential, and we note that the underlying first-order superlattices are visible. For clarity, we smoothed the data with a Gaussian filter of width $\sigma=3.5$~nm ($\sigma$ is the standard deviation of the Gaussian kernel). The scale bar is 250~nm.}
    \label{fig:formation-second-order}  
\end{figure*}

\vspace*{0.2cm}
\noindent
The recently observed collection of strongly correlated phases in magic-angle twisted bilayer graphene (MATBG) has sparked a wave of experimental and theoretical discoveries \cite{Cao2018_SC, Cao2018_corr, Yankowitz2019, Lu2019, Xie2020, po2018origin, dodaro2018phases, xie2020topology, Balents2020, lian2019twisted, gonzalez2019kohn, isobe2018unconventional, liu2018chiral, bistritzer2011moire, koshino2018maximally, kang2019strong, nuckolls2020strongly, Novelli2020, wu2021chern, saito2021hofstadter, saito2021isospin}. In these 2D heterostructures, the proximity effects of encapsulating layers to the MATBG plane can be precisely controlled and provide an additional tuning knob \cite{Stepanov2020, Arora2020}. By virtue of such proximity response, alignment of MATBG to adjacent layers of insulating hBN has a dramatic effect on the electronic properties \cite{long2022atomistic, Shi2021c, Cea2020a, shin2021electron, Mao2021a}, and even led to anomalous Hall phases in MATBG \cite{Stepanov2020, Arora2020} and ferroelectricity in bilayer graphene \cite{zheng2020unconventional}. \NT{Under specific alignment conditions, a second-order superlattice (SOSL) emerges as a result of geometrical interference between two underlying first-order moiré superlattices such as hBN-aligned graphene and MATBG superlattices (see Fig.~\ref{fig:intro-experiment}a and Fig.~\ref{fig:formation-second-order}a)} \cite{Wang2019, Finney2019, Wang2019_science, sinner2023strain, aggarwal2023moire}. Scanning tunneling spectroscopy experiments have showed the local imaging of such aligned heterostructures, yet only providing visualization and insights within a nanometer-size area\cite{oh2021evidence}. The real-space distribution of the SOSL potential on the mesoscale has not yet been reported, thus leaving many open questions about precursors for exotic quantum phases in graphene-based moiré heterostructures, and the role of structural characteristics that may stabilize these phases\cite{wagner2022global}. Furthermore, experimental investigations into the implications of superlattice strain have been limited, despite the fact that it’s always present in devices\cite{lau2022reproducibility}, and has a strong impact on the superlattice potential, resulting in a reconfiguration of the phase diagram\cite{wagner2022global, nuckolls2023quantum, uri2023superconductivity} and the emergence of novel quantum phases, such as stripe-like orders \cite{Jiang2019}.

In this work, we perform cryogenic near-field opto-electronic experiments on a MATBG aligned to hBN. With this technique, we are capable of probing the photovoltage response at length scales far below the Abbe diffraction limit, where we are only limited by the tip radius ($\approx 20$~nm) and by any spreading induced by the photoresponse mechanism. We observe two sets of fringes spatially rotated with respect to each other, which we interpret as a manifestation of large-scale local potential variations that originate from the second-order superlattice (SOSL). We complement our experimental findings with a theoretical model that visualizes the SOSL in real space as the interference between the underlying first-order superlattice potentials associated with the twisted bilayer graphene (TBG) and graphene/hBN superlattices. Our experiments, complemented by modelled potentials, show the high sensitivity of the resulting SOSL to local strain and twist angle variations. \NT{Finally, given the broken inversion symmetry that the SOSL implies, we discuss its impact on the flat-band physics in MATBG in the light of our novel PV nanoscopy data.}

Our experimental workflow is schematically pictured in Fig.~\ref{fig:intro-experiment}a. We fabricated a heterostructure consisting of hBN-encapsulated MATBG where the top encapsulating layer is closely aligned to the upper graphene layer as we verify with AFM measurements (Supplementary Note~\NTblue{I}). The heterostructure is contacted to a set of Cr/Au metallic leads (see Methods and Fig.~\ref{fig:intro-experiment}c). In order to visualize the SOSL potential, defined here as electronic potential landscape governed by the atomic lattice structure, we employ a scattering-type scanning near-field imaging microscope operating at temperature $T$ down to 10~K. By focusing infrared light (excitation energy $E=116$~meV, corresponding to the wavelength $\lambda_{IR}=10.7~\mu$m) to the apex of a sharp metal-coated AFM tip (radius of $\approx$ 20 nm), a hot-spot of light is generated inducing local photoexcitation of the charge carriers at the nanoscale. This leads to a local photovoltage generation that is probed by the global contacts, facilitated by the Shockley-Ramo mechanism \cite{ma2022photocurrent} (see Fig.~\ref{fig:intro-experiment}b,c). Details of the cryogenic near-field photovoltage measurements can be found in the Methods section.

\vspace*{0.2cm}
\noindent
\textbf{Photovoltage nanoscopy on a second-order superlattice} 
\newline
The main experimental observation is shown in Fig.~\ref{fig:intro-experiment}b, featuring a photovoltage map of our MATBG sample at $T=10$~K. Surprisingly, we observe photovoltage fringes that span across the entire bulk of the sample. In particular, in the top-left corner we observe a clear superposition of two sets of almost vertical and almost horizontal fringes (highlighted by the black circle). \NT{Supplementary Note~\NTblue{II} shows these maps for different contact pairs, from which we can conclude that these patterns cover the entire bulk of the sample.}

\begin{figure*}[t]
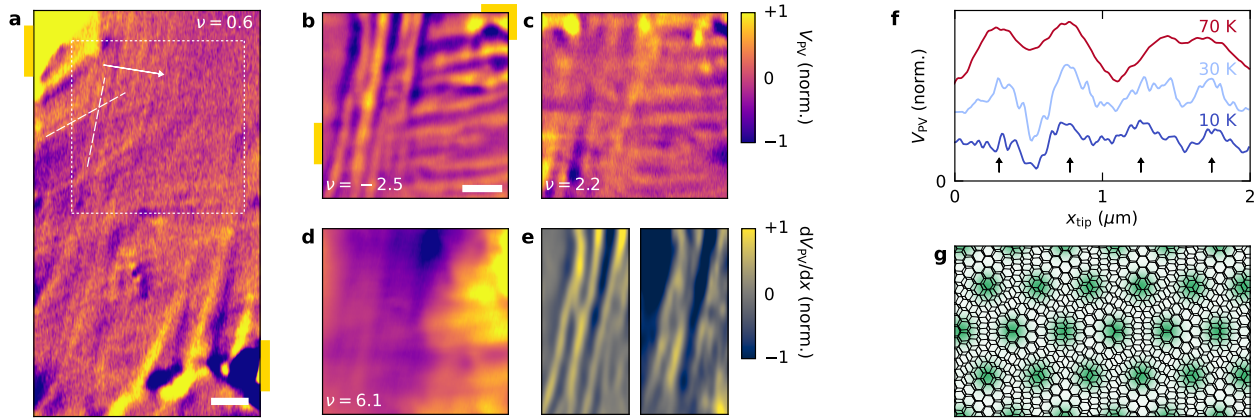

    \centering
    \includegraphics[scale=\scalefig]{\pathfigs /Fig2a-PV-B-10K.pdf}
    \includegraphics[scale=\scalefig]{\pathfigs /Fig2bcde-PV-doping.pdf}
    \includegraphics[scale=\scalefig]{\pathfigs /Fig2fg-PV-B-linetraces-inversion-symmetry-breaking.pdf}
    \caption{\textbf{Gate and temperature  response of the observed photovoltage features and broken inversion symmetry.} \textbf{a}, Local photovoltage map of the entire device taken at $\nu=0.6$ and $T=10$ K. The white dashed lines show two dominant directions of the observed fringes. The scale bar is 1~$\mu$m and the locations of the voltage probes are highlighted in yellow. \textbf{b}-\textbf{d}, Photovoltage maps taken at different filling factors in the area highlighted by the white dashed-line box in \textbf{a}. We observe no significant change in the overall fringe layout, even with the Fermi level in the remote bands (panel \textbf{d}). This excludes the effect of interactions being responsible for these periodic structures. \textbf{e} Derivatives ${\rm d}V_{\rm PV}/{\rm d}x$ taken for the left parts of the maps in \textbf{b} and \textbf{d} confirm the insensitivity of the underlying structure to the electronic state of TBG. \textbf{f} Photovoltage linecuts taken along the white arrow in \textbf{a} for three different temperatures. Positions of the common photovoltage peaks are highlighted by the black arrows, which remain the same over the presented temperatures range. \textbf{g}, Visualization of the broken inversion symmetry in a TBG structure. The lattice structure of TBG interferes with the triangular background potential owing to the hBN/graphene superlattice. Alltogether, inverting the structure around the central point would yield a different structure, and hence the inversion symmetry is broken.}
    \label{fig:doping-scans}  
\end{figure*}

To further unravel the nature of the observed photovoltage periodicities, we map the photovoltage response for different gate voltages (Fig.~\ref{fig:doping-scans}a-d). We observe a very similar photovoltage pattern at charge neutrality (Fig.~\ref{fig:doping-scans}a) and with the Fermi level inside the moiré flat bands (Fig.~\ref{fig:doping-scans}b and c), as well as at band fillings outside the flat band (Fig.~\ref{fig:doping-scans}d). In particular, the observed periodicity does not change with gate voltage. We also find that the photovoltage response is enhanced by an order of magnitude for fillings outside the flat band, with the features appearing to be less pronounced. (Fig.~\ref{fig:doping-scans}d). To accentuate that the underlying structure remains unchanged with the filling factor, we show in Fig.~\ref{fig:doping-scans}e the spatial derivatives of the photovoltage response and find a clear resemblance between the observed features at the Fermi levels inside and outside the flat-band fillings. As a consequence, the observed features cannot be linked with electronic interactions, because the electron dynamics change drastically between the charge densities inside and outside the flat-band fillings, but rather originate from a modulating potential background.

Additional evidence of a modulating potential background is further corroborated by photovoltage measurements as a function of temperature. Fig.~\ref{fig:doping-scans}f depicts normalized photovoltage line cuts and shows an unchanged response with temperature, aside from the magnitude (Supplementary Note~\NTblue{III} shows the full spatial photovoltage maps at different temperatures). This observation also suggests that the role of electronic interactions in creating the periodic photovoltage response remains irrelevant. From these line traces we extract a periodicity of the SOSL potential of $\sim 500 $~nm, which falls within the resolution limit of our experiment (see Supplementary Note~\NTblue{IV}).

\begin{figure}[!h]
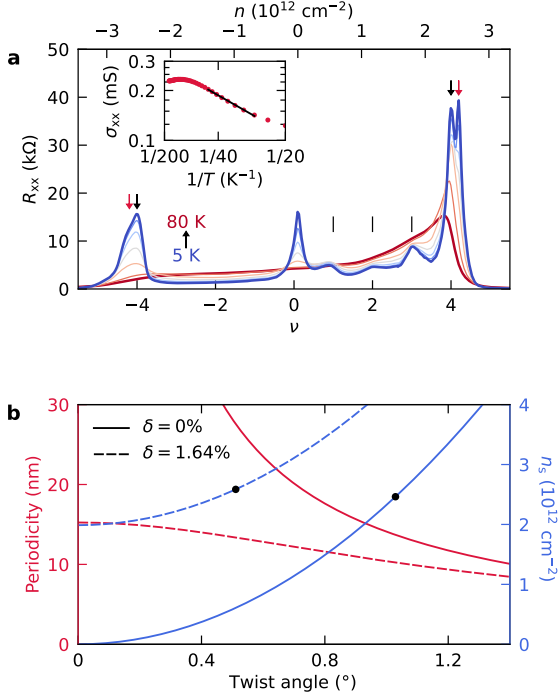

    \centering
    \includegraphics[scale=\scalefigtransport]{\pathfigs /Fig3a-transport.pdf}
    \includegraphics[scale=\scalefigtransport]{\pathfigs /Fig3b-full-filling-periodicity.pdf}
    \caption{\textbf{Electronic transport measurements.} \textbf{a}, Longitudinal resistivity measurements as a function of MATBG filling factor at different temperatures ranging from 5 to 80 K. Solid black lines indicate positions of the integer fillings $\nu=1$, 2, 3. Black arrows correspond to the resistance peaks originating from fully-filled ($\nu=+4$) and fully-emptied ($\nu=-4$) MATBG moiré superlattice. Red arrows denote the hBN-induced energy gaps. \NT{Inset: Arrhenius plot of the longitudinal conductivity at the CNP. A fit (see Methods) yields a thermally activated gap size of 3.6~meV.} \textbf{b}, Theoretical calculations of the superlattice periodicity (red) and corresponding fully filled moiré charge carrier density (blue) as a function of the twist angle $\theta_{\rm TBG}$ (solid lines) and $\theta_{\rm hBN}$ (dashed lines). Here, we assume a lattice constant mismatch between graphene and hBN of $\delta=1.64$\%. The black dots correspond to the peaks in \textbf{a} marked by the arrows, from which we deduce twist angles $\theta_{\rm TBG} = 1.03 \degree$ and $\theta_{\rm hBN} = 0.51 \degree$.}
    \label{fig:transport}  
\end{figure}

In order to confirm the presence of a SOSL and further characterize it, we explore its low-temperature electronic transport properties (see Methods for details). \NT{We utilize these measurements to probe the details of the underlying electronic band structure of MATBG aligned to hBN. In particular, the flat-band physics clearly reveals itself within the filling factors $\nu = \pm 4$, corresponding to the fully-emptied or -filled MATBG superlattice unit cell.} In Fig.~\ref{fig:transport}a, a set of transport curves $R_{xx}$ vs $\nu$ taken at temperatures between 5~K and 80~K indicates the presence of flat bands in our twisted bilayer graphene device. We identify characteristic local resistance maxima at integer fillings close to $\nu$=+1, +2, and +3 (marked by black lines), confirming the magic-angle nature of our device. Our transport measurements show two pronounced band insulator resistance peaks characteristic for fully-filled or -emptied electronic flat bands at $\nu = \pm 4$. Interestingly, instead of observing a single resistance peak, we observe split resistance maxima at $\nu = +4$ and a shoulder at $\nu = -4$ that appear nearly symmetric around the charge-neutrality point (CNP). Here, we argue that this electronic transport signature originates from close alignment between hBN and MATBG. \NT{Our presumption is mainly supported by the presence of a large thermally-activated gap at CNP of $\Delta_{\mathrm{CNP}} = 3.6$~meV (see inset Fig.~\ref{fig:transport}a). This gap signals an opening of the Dirac cones due to the graphene/hBN alignment\cite{Zhang2019}, and the extracted gap size is similar to the values found in other hBN-aligned MATBG devices studied elsewhere\cite{Serlin2020, Sharpe2019}. The hBN/graphene alignment is further validated by inspecting the AFM topography (Supplementary Note~\NTblue{I}). With this in mind, we conclude that the lower density peak in Fig.~\ref{fig:transport}a originates from a single-particle gap in the MATBG spectrum (marked by the black arrow), while the higher density peak is a result of the alignment between the MATBG and hBN planes (marked by the red arrow) \cite{Dean2013, Yankowitz2012}.}

After carefully extracting the charge carrier density using the Hall effect, we define the twist angles $\theta_{\rm hBN}$ (between the hBN and the bottom graphene) and $\theta_{\rm TBG}$ (between the two graphene layers) based on the resistance peak positions (see Methods). Fig.~\ref{fig:transport}b relates the charge carrier density expected at a fully filled moiré band in the case of the twisted graphene layers (solid lines, lattice mismatch $\delta = 0 \%$) and in the case of graphene/hBN (dashed lines, lattice mismatch $\delta = 1.64 \%$ \cite{Cao2018_corr}). Here, the black markers correspond to the charge carrier densities extracted from Fig.~\ref{fig:transport}a, from where we determine $\theta_{\rm TBG} = 1.03 \degree$ and $\theta_{\rm hBN} = 0.51\degree$. We note that our device shows a small twist angle variations of $\pm 0.02 \degree$ (Supplementary Note~\NTblue{V}), in line with the presence of a SOSL across the entire sample, despite its high twist-angle sensitivity. \NT{From these transport measurements we deduce that the sample hosts two coexisting superlattices, formed by a graphene layer closely aligned to hBN and the TBG superlattice itself. Their periodicities (deduced as in Fig.~\ref{fig:transport}b) of $\lambda_{\rm M}$ of $\sim$13.4 and $\sim$13.7~nm, respectively, are sufficiently close to give rise to a SOSL with an even larger periodicity $\tilde{\lambda}_{\rm M}$ \cite{Wang2019, Finney2019, Wang2019_science}. Fig.~\ref{fig:intro-experiment}a shows an example of such a second-order superlattice. To strengthen our findings, we also performed measurements on a MATBG device without close alignment between hBN and TBG, and found that all the signatures related to the SOSL are absent in the spatially-resolved $PV$ maps (Supplementary Note~\NTblue{VI}).}

\vspace*{0.2cm}
\noindent
\textbf{Calculation of spatial potential profile}
\newline
To model the spatial potential profile of the second-order superlattice, we examine its formation in the reciprocal space. In this work, we consider the case of an unstrained and strained heterostructure. First, we start with the case without applied strain. Figure~\ref{fig:formation-second-order}a shows the lattice vectors $\vec{k}$ of the unstrained hBN and graphene lattices, from which we take the difference vectors (blue) that represent their SOSL. These vectors feature a strongly reduced length and by taking the inverse of their moduli (and accounting for a factor $2 \pi$), we find the periodicity $\tilde{\lambda}_{\rm M}$ of the SOSL (also defined analytically in the Methods). Figure~\ref{fig:formation-second-order}c depicts the SOSL periodicity for a range of $\theta_{\rm hBN}$ and $\theta_{\rm TBG}$. We note that $\tilde{\lambda}_{\rm M}$ is equal in the three principal directions (representing a triangular lattice). We find a limited window of twist angles where a SOSL can be detected in our experiment, that is where $\tilde{\lambda}_{\rm M} \gg \lambda_{\rm M}$. This is understood as in reciprocal space the superlattice vectors need to line up very closely with each other, while they also depend sensitively on the hBN and graphene lattice mismatch $\delta$, $\theta _{\rm hBN}$, and $\theta _{\rm TBG}$. For $\delta=1.64 \%$ and the twist angles extracted from the transport measurements, we find $\tilde{\lambda} _{\rm M} \approx 390$~nm, which is slightly lower than the periodicity extracted in the near-field photovoltage maps. Note, it is a coincidence that the hBN-graphene lattice mismatch occurs such that SOSLs appear for twist angles near the magic angle. 

\begin{figure*}[t]
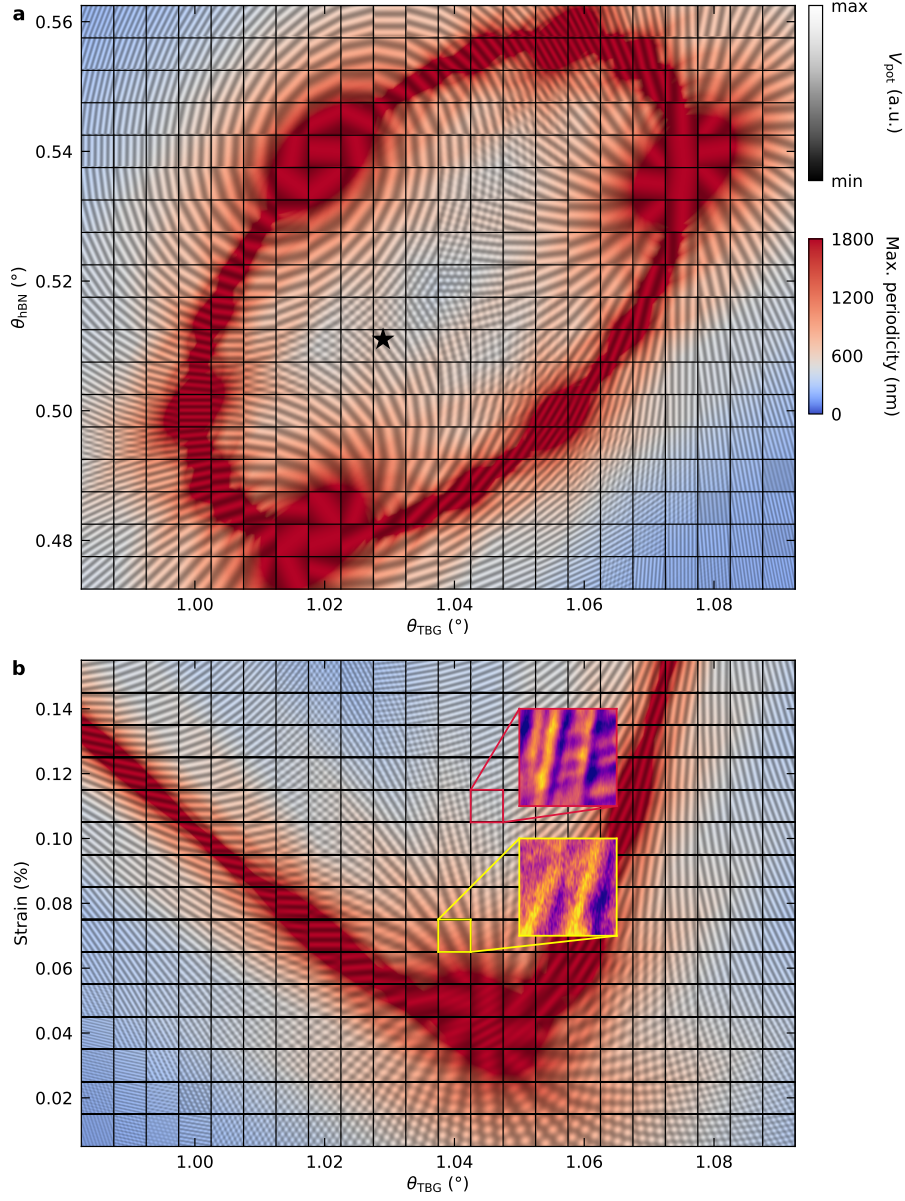

    \centering
    \includegraphics[scale=\scalefigfive]{\pathfigs /Fig5a-lattice-birdview-Lcool-180nm-1800nm-5x-1200px.pdf}
    \includegraphics[scale=\scalefigfive]{\pathfigs /Fig5b-Lcool-180nm-1800nm-5x-1200px-strain.pdf}
    \caption{\textbf{Real-space maps of the second-order superlattice as a function of twist angle and strain magnitude.} \textbf{a} In each square, the grey-scaled color map represents a calculation of the total superlattice potential across an area of $1800\times 1800$~nm$^2$ with $\theta_{\rm TBG}$ and $\theta_{\rm hBN}$ given by the center of the square. The applied strain is the same as that in the other figures (0.1\% along the zigzag direction). To simulate the effects of the finite tip radius and cooling length, we smoothed the data with a Gaussian filter of width $\sigma=180$~nm ($\sigma$ is the standard deviation of the Gaussian kernel). The colored background is the same calculation of the maximum periodicity along one of the SOSL direction as presented in Fig.~\ref{fig:formation-second-order}\textbf{d}. \textbf{b} Same as \textbf{a} but as function of strain magnitude with $\theta_{\rm hBN}$ fixed at $0.51 \degree$. The two insets are the experimental data \NT{for the contact configuration as shown in Fig.~\ref{fig:doping-scans}\textbf{a} and Fig.~\ref{fig:intro-experiment}b}, which were found to be qualitatively consistent with the highlighted calculated real-space maps. The area of the insets is also $1800\times 1800$~nm$^2$.}
    \label{fig:lattice-birdview}  
\end{figure*}

We now move to the case where the lattice strains become important actors in the overall SOSL formation. Our minimalistic model of a triangular lattice lacks an explanation for the observation of periodic fringes in only two directions, which in turn are separated by an unexpected angle of $\sim 50 \degree$ (dashed lines in Fig.~\ref{fig:doping-scans}a). To account for that, we introduce a small amount of a uniaxial heterostrain $\epsilon$ to the graphene layers, which is commonly present in these heterostructures \cite{wong2020cascade, Jiang2020, Benschop2021, Kazmierczak2021, Mesple2020}. We apply a strain tensor $\overline{\epsilon}$ to the graphene lattice vectors \cite{Pereira2009, Naumis2017}, following $(\overline{I} + \overline{\epsilon})^{-1} \vec{k}$, where
\begin{equation}
\label{eq:strain-tensor}
\overline{\epsilon} = \epsilon
\begin{pmatrix}
\cos ^2 \alpha  - \rho \sin ^2 \alpha   & (1+\rho) \cos \alpha \sin \alpha \\
(1+\rho) \cos \alpha \sin \alpha        & \sin ^2 \alpha  - \rho \cos ^2 \alpha
\end{pmatrix} ,
\end{equation}
where $\alpha$ is the angle of the principal direction of the applied uniaxial strain with respect to the zigzag direction, and $\rho = 0.165$ is the Poisson ratio of graphene \cite{Blakslee1970}. $\overline{I}$ is the identity matrix. As illustrated in Fig.~\ref{fig:formation-second-order}b, strain has a large impact on the relative magnitude and direction of the second-order superlattice vectors. This leads to a deformation of the lattice, such that the superlattices act as `magnifying glasses' of strain \cite{Cosma2014, Jiang2017_strain}. We repeat the calculation of $\tilde{\lambda} _{\rm M}$ with $0.1 \%$ strain\cite{wong2020cascade, Jiang2020, Benschop2021, Kazmierczak2021, Mesple2020} applied to both graphene layers along the zigzag direction (Supplementary Note~\NTblue{VII} discusses details of these calculations in the presence of strain). The blue vectors in Fig.~\ref{fig:formation-second-order}b become modified and now have three different moduli, which would naively be represented by three separate red color blobs in the $\theta_{\rm TBG}$-$\theta_{\rm hBN}$ space. 

Figure~\ref{fig:formation-second-order}d displays the maximum periodicity of three principal lattice vectors and reveals an intricate picture of strongly varying periodicities for different twist angles. Notably, this distribution is much more complex than expected from the simple picture introduced in Fig.~\ref{fig:formation-second-order}a. Here, we take into account that a simple vector summation may end up with a set of lattice vectors representing by obtuse triangles. Since that would lead to an overestimation of $\tilde{\lambda} _{\rm M}$, we correct for it by remapping the lattice vectors until we reach an acute triangle, allowing for a fair comparison of $\tilde{\lambda} _{\rm M}$ with $\lambda _{\rm M}$. This leads to the SOSL unit cell size that is represented by a fractal-like ring of non-vanishing wavelengths (Fig.~\ref{fig:formation-second-order}d). \NT{Based on the observed and calculated SOSL periodicities, our phenomenological model characterizes the experimental observations very well without utilizing any fit parameters. At the same time, we highlight that the used set of parameters is not unique and one can identify complimentary sets that would also account for the presented data. Supplementary Note~\NTblue{VIII} discusses the effect of different hBN lattice mismatch, stain magnitude and strain angle.}

To obtain a better understanding of this result, we focus our attention to three particular sets of parameters ($\theta _{\rm hBN}$, $\theta _{\rm TBG}$, $\epsilon$) and visualize the real-space potential of the resulting second-order superlattice \NT{(method explained in Supplementary Note~\NTblue{IX})}. As shown in Fig.~\ref{fig:formation-second-order}e, extremely small variations in the twist angle of order 0.01$\degree$ and strain variations of 0.01\% can have a large impact on the type of second-order superlattice. For instance, by applying 0.1\% of strain, the geometry may change from a triangular to 1D lattice \cite{sinner2023strain}. Likewise, a square lattice can be formed by changing both twist angles only slightly. We calculate the real-space potential for a wider range of both twist angles, as presented in the birds-eye view in Figure~\ref{fig:lattice-birdview}a. We emphasize that smaller periodicities must emerge in other non-dominant directions as well. However, these cannot be measured using our near-field probe with a limited spatial resolution and spatial selectivity of the global electrical probes. 

We can identify the following features: at three resonant points, a 1D~lattice emerges with an even longer periodicity than our sample lateral dimensions, and hence, is not detectable. Surrounding these points, 1D~lattice structures are present with periodicities on the order of several hundreds of nanometers. Further away from these resonant points, the periodicity tends to fall below the experimental resolution, and thus, could not be observed. On the other hand, a triangular lattice only forms right in the middle of these resonant points, which in turn converts via a square lattice towards a 1D lattice near these resonant points. This clearly illustrates the wide variety of lattices that can be formed in the very limited parameter space of our experiment: ($\theta _{\rm hBN}$, $\theta _{\rm TBG}$, $\epsilon$). \NT{Based on the distribution of lattice types, it is not surprising that the triangular lattice is not detected in our measurements. The triangular lattice appears only in a very narrow region in the center of Fig.~\ref{fig:lattice-birdview}a surrounded by the regions of rectangular lattices. The remainder of the parameter space is dominated by the 1D type. Here, we also note that our observations are expected to be ubiquitous in moiré heterostructures, and Supplementary Note~\NTblue{X} provides theoretical modelling for the case of twisted trilayer graphene.}

To emphasize the effect of strain fields, Fig.~\ref{fig:lattice-birdview}b shows another perspective of the lattice types as a function of strain along the zigzag direction and the twist angle in the graphene-graphene superlattice with $\theta _{\rm hBN}$ fixed at $0.51\degree$. Taking the experimental value for $\theta _{\rm TBG} = 1.03 \pm 0.02 \degree$, we identify locations on this map that closely resemble our PV data. The red inset shows a square-like SOSL at $\theta _{\rm TBG} = 1.045\degree$ and $\epsilon = 0.11\%$ that is consistent with the upper-left quadrant of Fig.~\ref{fig:intro-experiment}b. In addition, the yellow inset shows a 1D set of PV fringes taken from the bottom-right corner of Fig.~\ref{fig:doping-scans}a, and is similar to the calculated SOSL at $\theta _{\rm TBG} = 1.04\degree$ and $\epsilon = 0.07\%$. This highlights that changes of scale 0.01\% in $\epsilon$ can result in a significant change of the type of SOSLs in the same sample providing us a pathway to manipulate SOSL \textit{in situ}. 

We note that the spatial photovoltage profile as measured by the photovoltage nanoscopy reflects in certain cases only partially the underlying SOSL structure. This is inevitable as photovoltage nanoscopy is typically sensitive to the projection of the gradient of the Seebeck coefficient onto the local current flow \cite{ma2022photocurrent}. To ensure that we have a complete picture of the underlying SOSL structure, we perform theoretical simulations of the expected PV maps for three different configurations of the contact probes (Supplementary Note~\NTblue{XI}). Comparing them with the experimental data, we find that a major part of our sample features a square SOSL (as shown in Fig.~\ref{fig:lattice-birdview}b).

\vspace*{0.2cm}
\noindent
\NT{
\textbf{Impact of broken inversion symmetry on flat-band physics}
\newline
Our photovoltage maps provide a real-space image of the broken inversion symmetry in the MATBG superlattice for the entire bulk of the studied sample. MATBG becomes asymmetric with respect to inversion when closely aligned to the adjacent hBN layer (Fig.~\ref{fig:doping-scans}g). To what extent can the presence of the SOSL affect the physics of MATBG?

Our device exhibits correlated states, as indicated by the resistance peaks at integer fillings (Fig.~\ref{fig:transport}a). At first glance, it appears surprising that the magic angle physics persists in this sample, given that the macroscopic signature of SOSL extends across the entire device. Yet, it has been shown that such underlying lattice potential does not have a destructive effect on the flat-band physics \cite{Serlin2020, long2022atomistic}. Nonetheless, when we have a more in-depth look at our data, we do observe some profound effects of the superlattice that perturb the physics of MATBG.  

For the hole doping, the flat band is shifting closer in energy to the remote valence band, as evidenced by the relatively small gap of 13 meV extracted from the transport measurements (Supplementary Note~\NTblue{V}). Consequently, from a theoretical \cite{Xie2020} and experimental \cite{oh2021evidence} perspective, this diminishes the chance of observing strongly-correlated phases and superconductivity since typically well-defined gaps and an isolated flat band are required. Indeed, as seen in our data, the correlations are suppressed on the hole side (Fig.~\ref{fig:transport}a), which marks a strong asymmetry between the hole- and electron-like flat bands. In addition, given the broken inversion symmetry, we do not observe any signatures of the anomalous Hall effect (AHE) in our electronic transport measurements. This is consistent with the recent theoretical predictions and local magnetometry experiments, where only very specific twist angles between MATBG and hBN/graphene satisfy a commensurate condition for the AHE state \cite{shi2021moire, grover2022chern}. 

In addition, the broken inversion symmetry strongly influences the second-order electrical response in these moiré systems \cite{ma2022photocurrent}. This symmetry breaking induces a non-trivial quantum geometry, offering an exciting route towards probing the interplay of quantum geometry and strong correlations. In particular, this enables the generation of the so-called shift currents \cite{kaplan2022twisted}. As part of ongoing work on the same device \cite{inpreparation}, we observe indications of a shift-current generation as well as signatures of a global cascade of phase transitions similar to those reported in local spectroscopy experiments \cite{wong2020cascade, Zondiner2020}.}

\vspace*{0.2 cm}
\noindent
\textbf{Outlook} 
\newline
To conclude, we have realized and observed a second-order superlattice formed by the alignment of MATBG to one of the adjacent hBN substrates. These observations, corroborated by our theoretical model, show an utmost tuneability of the SOSL structures. In combination with controlled ways \cite{Ribeiro-Palau2018, inbar2022quantum} to tune $\theta_{\rm hBN}$, $\theta_{\rm TBG}$ or strain $\epsilon$, it will open the pathway to explore optoelectronic properties of a variety of different SOSL structures including triangular, square and 1D lattices. In particular, the latter holds a promise to explore Luttinger liquid states with a tuneable crystalline quality 1D channels\cite{wang2022one, sinner2023strain}. The alignment of MATBG and hBN has been found to promote the emergence of unique quantum states displaying the quantum anomalous Hall effect\cite{tschirhart2020}. Our study sheds a new light on the mesoscale precursors of such alignment. Serving as a magnifying glass to reveal real-space strain fields, SOSLs open a pathway to realize quality control in MATBG devices and thus lead to a more consistent fabrication process of moiré samples\cite{lau2022reproducibility}. Some open questions, however, remain. For example, what exact mechanism drives the photovoltage in the SOSLs and what is at the cause of the required non-linear effect playing a significant role in the photovoltage generation? As explored in Supplementary Note~\NTblue{XII}, the photothermoelectric (PTE) effect and the second-order photoresponse are the two main candidates that could generate the observed photoresponse. While the PTE is typically present in graphene-based devices and resembles key signatures of our observations such as the doping- and temperature dependence, the second-order photoresponse has also been demonstrated to produce a strong infrared photoresponse because of the non-zero Berry curvature native to MATBG's electronic flat bands \cite{ma2022intelligent}. Further experiments are needed to reveal a mesoscopic picture of inversion symmetry breaking in moiré materials due to alignment with hBN and the role of strain profiles.

\onecolumngrid
\vspace*{0.5 cm}
\hrule
\vspace*{0.2 cm}
\twocolumngrid

\section*{\textsf{Methods}}
\small
\subsection*{Device fabrication}
\noindent
The device consists of TBG encapsulated in 16~nm bottom hBN and 10~nm top hBN flakes, altogether placed on top of a graphite flake, serving as a local gate. During the stacking process, the graphene flake is cut with an AFM tip, with intention to prevent additional strain building up in the tear-and-stack process used otherwise \cite{Chen2019, Saito2020, Stepanov2020}. To minimize the number of air bubbles in the stack, we pick up each flake at a temperature within $100-110$~$\degree$C \cite{Pizzocchero2016, Purdie2018}. In the final step, when dropping the stack on the target substrate with alignment markers, we repeat the drop-down step at least once to further squeeze out air bubbles. Figure~\NTblue{S1} shows an AFM scan of the resulting stack. We choose the cleanest area of the stack to pattern our device in a Hall-bar shape (Fig.~\ref{fig:intro-experiment}c).

\subsection*{Cryogenic near-field photovoltage measurement details}
\noindent
We used a cryogenic scattering-type scanning near-field microscope (cryoSNOM) developed by Neaspec/Attocube to carry our the near-field photovoltage experiments at temperatures between $10-300$~K. A tuneable quantum cascade laser (Daylight Solutions) acts as an infrared light source, and the data shown in this work were acquired at an excitation energy of $116$~meV (10.6~$\mu$m). We focus approximately 10~mW of this light on a PtIr-coated AFM tip (Nanoworld, 50~nm coating), which is oscillating above the sample surface at $\approx 250$~kHz with a tapping amplitude of $\approx 100$~nm. The AFM feedback loop incorporates a system developed by Neaspec/Attocube by which we can lower the quality factor of the AFM cantilever resonance to the values similar to ambient operation at room temperature. This helps a quick decay of the cantilever motion, and therefore we are less limited in the scanning speed. Finally, to reduce coupling of strong floor vibrations with our microscope, we set up a home-built active damping system that cancels these vibrations and stabilizes the optical table.

For simultaneous measurement of the photovoltage between two pairs of contacts, we used two differential voltage amplifiers (Ithaco 1201) with a different contact providing the ground. The carrier doping in our samples is tuned by applying a DC voltage between the graphite gate and our device, while keeping the $Si^{++}$ backgate grounded. To avoid detecting unwanted far-field contributions to the photovoltage signal, we detect the near-field signals at the second harmonic of the cantilever oscillation. \NT{In addition, we acquired the optically scattered near-field signal, but found that it is not sensitive to the features of the SOSL.}

We follow Ref.~\citenum{Hesp2021} in the scheme for analysing the photovoltage maps. Here, the measured photovoltage signal is demodulated with the driving signal of the AFM cantilever as a reference signal. However, the actual motion of the AFM cantilever can have a phase offset that varies with the position on the sample (due to tip–sample interaction). This phase offset is given at each pixel by the measured phase delay between the tip driving signal and the actually detected motion. Therefore, we correct our photovoltage signal measured at harmonic $i$ by subtracting at every point $i$ times this phase delay. In addition to this, there remains a global phase offset in the corrected photovoltage signal due to the electronics in the circuit. Since the photovoltage signal is a real-valued quantity, we subtract this global phase offset, which we determine by taking the most frequent phase within a scan. In this work, we use the second harmonic ($i=2$) of the photovoltage signals.

\subsection*{Transport measurement details}
\noindent
The main set of four-terminal transport data shown in Fig.~\ref{fig:transport}a was taken in an Advanced Research System cryostat with the base temperature of 5 K and the magnetic field up to 1 T. In these electronic transport measurements we followed a conventional lock-in measurement scheme. A low-frequency AC current (17.111~Hz) of 10 nA flows between the bottom-right and top contacts, while measuring the voltage drop between the two left-middle contacts using a Stanford SR860 lock-in amplifier. The gate voltage was sourced using a Keithley 2400 Source Meter Unit.

\subsection*{Analysis of transport data}
\noindent
From a Hall measurement in a magnetic field of 1~T we determine the carrier density $n$ (in cm$^{-2}$) as a function of the applied gate voltage $V_{\rm G}$. A linear fit near the charge-neutrality points ($|V_{\rm G}|< 1$~V) yields $n(V_{\rm G}) = 1.24 \cdot V_{\rm G} + 0.27$ in units of $10^{12}$~cm$^{-2}$. For the subsequent analysis we allow for a small shift in the charge neutrality voltage (for instance induced by photodoping \cite{Ju2014}) by replacing $V_{\rm G}$ with $V_{\rm G} - V_{\rm shift}$. From the peak value of the resistance peak at charge neutrality in Fig.~\ref{fig:transport}a we find $V_{\rm shift} = -45$~mV. We define the full-filling carrier density $n_{\rm s}$ as half the distance from the two resistance peaks marked by the black arrows in Fig.~\ref{fig:transport}a. In this, we notice that these two peaks are slightly off-centered with respect to the charge neutrality by 60~mV. The resistance peak of the hBN-Gr superlattice is less developed for negative carrier densities. Therefore we extract the $n_{\rm s}$ corresponding to the hBN-Gr lattice from the relative scaling of the Gr-Gr and hBN-Gr superlattice peaks at positive carrier densities (marked in black and red, respectively). \NT{We obtain the band filling factor as $\nu = 4n/n_s$ (accounting for spin/valley degeneracy). The thermal activation of a band gap allows us to determine the gap size $\Delta$ by fitting our data to $\sigma_{\rm xx} \propto e^{\Delta/2k_{\rm B}T}$.}

\subsection*{First-order superlattice periodicity}
\noindent
Our transport data reveals two moiré lattices hosted by our sample: one given by graphene aligned to hBN, and another one owing to TBG. Such moiré lattices exhibit a resistive state at a particular carrier density associated with the full-fulling state of the superlattice \cite{Yankowitz2012, Cao2016}. For a superlattice formed by two superposed hexagonal lattices, this full-filling carrier density $n_{\rm s}$ is given by \cite{Yankowitz2012}
\begin{equation}
\label{eq:superlattice-full-filling}
n_{\rm s} = \frac{8}{\sqrt{3} \lambda_{\rm M} ^2} , 
\end{equation}
where the twist angle $\theta$ and lattice mismatch $\delta$ define the superlattice periodicity $\lambda _{\rm M}$ as
\begin{equation}
\label{eq:superlattice-periodicity}
\lambda _{\rm M} = \frac{(1+\delta) a}{\sqrt{2 (1+\delta)(1-\cos \theta) + \delta^2}}
\end{equation}
with $a=0.246$~nm corresponding to the graphene lattice periodicity. The properties of TBG are reflected by the curves with $\delta = 0 \%$, while we model hBN-graphene superlattice with $\delta = 1.64  \%$. This value is slightly lower than the value typically stated in literature, $\delta =1.8 \%$, However, the typical amount of strain in the graphene layers of $\approx 0.1 \%$ can account for this difference.

\bibliographystyle{naturemag-achim}
\bibliography{main.bib}

\vspace*{-0.2cm}
\section*{\textsf{Acknowledgements}}
\vspace*{-0.2cm}
{\small
\noindent
F.H.L.K. acknowledges financial support from the Government of Catalonia trough the SGR grant, and from the Spanish Ministry of Economy and Competitiveness, through the Severo Ochoa Programme for Centres of Excellence in R\&D (Ref. SEV-2015-0522), and Explora Ciencia  (Ref. FIS2017-91599-EXP). F.H.L.K. also acknowledges support by Fundacio Cellex Barcelona, Generalitat de Catalunya through the CERCA program, and the Mineco grant Plan Nacional (Ref. FIS2016-81044-P) and the Agency for Management of University and Research Grants (AGAUR) (Ref. 2017-SGR-1656).  Furthermore, the research leading to these results has received funding from the European Union's Horizon 2020 programme under grant agreements Refs. 785219 (Graphene Flagship Core2) and 881603 (Graphene Flagship Core3), and Ref. 820378 (Quantum Flagship). This work was supported by the ERC under grant agreement Ref. 726001 (TOPONANOP). P.S. acknowledges support from the European Union’s Horizon 2020 research and innovation programme under the Marie Skłodowska-Curie Grant No. 754510.
N.C.H.H. acknowledges funding from the European Union's Horizon 2020 research and innovation programme under the Marie Skłodowska-Curie grant agreement Ref. 665884.
K.W. and T.T. acknowledge support from JSPS KAKENHI (Grant Refs. 19H05790, 20H00354 and 21H05233).
}
This project has received funding from the “Presidencia de la Agencia Estatal de Investigación” within the PRE2020-094404 predoctoral fellowship.

\vspace*{-0.2cm}
\section*{\textsf{Author contributions}}
\vspace*{-0.2cm}
{\small
\noindent
F.H.L.K. conceived the experiment. 
N.C.H.H., S.B.-P., P.S. performed near-field experiments on a system optimized by N.C.H.H., D.B.-R. and H.H.S.
Transport experiments were performed by P.S. on a system built by R.K.K. 
The sample was fabricated by P.S. using a contact recipe developed by H.A. and with hBN crystals provided by K.W. and T.T.
The results were analyzed and interpreted by N.C.H.H. and P.S. using a model developed by N.C.H.H. 
The manuscript was written by P.S., N.C.H.H. and F.H.L.K. with input from all authors. 
F.H.L.K supervised the work.
}

\vspace*{0.6cm}
\section*{\textsf{Competing Financial Interests}}
\vspace*{-0.2cm}
{\small
\noindent
The authors declare no competing financial interests.
}

\vspace*{-0.2cm}
\section*{\textsf{Data Availability Statement}}
\vspace*{-0.2cm}
{\small
\noindent
The data that support the plots within this paper and other findings of this study are available from the corresponding authors upon reasonable request.
}

\end{document}


\title{\Large\textsf{\papertitle}}

\author{Niels C.H. Hesp}
\affiliation{\footnotesize ICFO-Institut de Ci\`{e}ncies Fot\`{o}niques, The Barcelona Institute of Science and Technology, Av. Carl Friedrich Gauss 3, 08860 Castelldefels (Barcelona),~Spain}

\author{Sergi Batlle-Porro}
\affiliation{\footnotesize ICFO-Institut de Ci\`{e}ncies Fot\`{o}niques, The Barcelona Institute of Science and Technology, Av. Carl Friedrich Gauss 3, 08860 Castelldefels (Barcelona),~Spain}

\author{Roshan Krishna Kumar}
\affiliation{\footnotesize ICFO-Institut de Ci\`{e}ncies Fot\`{o}niques, The Barcelona Institute of Science and Technology, Av. Carl Friedrich Gauss 3, 08860 Castelldefels (Barcelona),~Spain}

\author{Hitesh Agarwal}
\affiliation{\footnotesize ICFO-Institut de Ci\`{e}ncies Fot\`{o}niques, The Barcelona Institute of Science and Technology, Av. Carl Friedrich Gauss 3, 08860 Castelldefels (Barcelona),~Spain}

\author{David Barcons-Ruiz}
\affiliation{\footnotesize ICFO-Institut de Ci\`{e}ncies Fot\`{o}niques, The Barcelona Institute of Science and Technology, Av. Carl Friedrich Gauss 3, 08860 Castelldefels (Barcelona),~Spain}

\author{Hanan Herzig Sheinfux}
\affiliation{\footnotesize ICFO-Institut de Ci\`{e}ncies Fot\`{o}niques, The Barcelona Institute of Science and Technology, Av. Carl Friedrich Gauss 3, 08860 Castelldefels (Barcelona),~Spain}

\author{Kenji Watanabe}
\affiliation{\footnotesize Research Center for Functional Materials, National Institute for Materials Science, 1-1 Namiki, Tsukuba 305-0044,~Japan}

\author{Takashi Taniguchi}
\affiliation{\footnotesize International Center for Materials Nanoarchitectonics, National Institute for Materials Science,  1-1 Namiki, Tsukuba 305-0044,~Japan}

\author{Petr Stepanov}
\email{pstepano@nd.edu}
\affiliation{\footnotesize ICFO-Institut de Ci\`{e}ncies Fot\`{o}niques, The Barcelona Institute of Science and Technology, Av. Carl Friedrich Gauss 3, 08860 Castelldefels (Barcelona),~Spain}
\affiliation{\footnotesize Department of Physics and Astronomy, University of Notre Dame, Notre Dame, IN 46556,~USA}
\affiliation{\footnotesize Stavropoulos Center for Complex Quantum Matter, University of Notre Dame, Notre Dame, IN 46556,~USA}

\author{Frank H.L. Koppens}
\email{frank.koppens@icfo.eu}
\affiliation{\footnotesize ICFO-Institut de Ci\`{e}ncies Fot\`{o}niques, The Barcelona Institute of Science and Technology, Av. Carl Friedrich Gauss 3, 08860 Castelldefels (Barcelona),~Spain}
\affiliation{\footnotesize ICREA-Instituci\'{o} Catalana de Recerca i Estudis Avan\c{c}ats, 08010 Barcelona, Spain}

\maketitle

\def\bibsection{\section*{\textsf{Supplementary references}}} 

\renewcommand{\figurename}{Fig.}
\setcounter{equation}{0}
\setcounter{figure}{0}
\setcounter{table}{0}
\makeatletter
\renewcommand{\theequation}{S\arabic{equation}}
\renewcommand{\thefigure}{S\arabic{figure}}
\renewcommand{\thetable}{\arabic{table}}
\renewcommand{\bibnumfmt}[1]{[#1]}
\renewcommand{\citenumfont}[1]{#1}

\onecolumngrid

\section{Estimation of twist angles based on AFM}
\label{sec:app-afm}
\begin{figure}[h]
    \centering
    \includegraphics{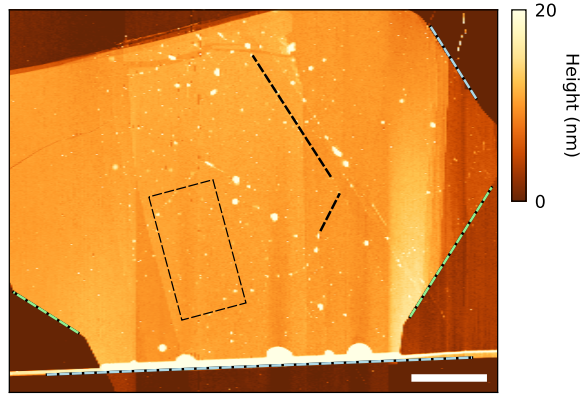}
    \caption{\textbf{Close alignment of graphene and hBN sheets in our device.} AFM image of our stack with the relative alignment of the upper graphene sheet (black), top hBN flake (blue) and bottom hBN layer (green). We extract the angles of $\approx 0.5 \degree$ ($5.0 \degree$) between the crystallographic axes of the top (bottom) hBN sheet and upper graphene layer. The rectangle marks the region where the Hall-bar is patterned and the scale bar is 8~$\mu$m.}
    \label{fig:app-afm}  
\end{figure}

\noindent
We can obtain an estimate of the relative alignment between the graphene and hBN sheets by comparing their crystallographic axes. Each flake features multiple straight edges in the AFM scan (Fig.~\ref{fig:app-afm}), which correspond either to the zigzag or armchair crystallographic direction. Based on these two types of edges for each flake, we find that the upper graphene sheet is misaligned to the top (bottom) hBN flake by $\approx 0.5 \degree$ ($5.0 \degree$), respectively. Since we cannot distinguish between zigzag and armchair directions, there is a probability of $50 \%$ that the layers are actually in close alignment, rather than being rotated by $\sim 30 \degree$. Raman spectroscopy is capable of addressing this ambiguity \cite{Eckmann2013}. We note that the relative orientations between the flakes can change slightly during the subsequent nano-fabrication steps. However, the following measurements indicate that this did not happen.

\section{Imaging second-order superlattice with different contact pairs} 
\label{sec:app-different-contacts}
\begin{figure*}[t!]
    \centering
    \includegraphics[scale=\scalefig]{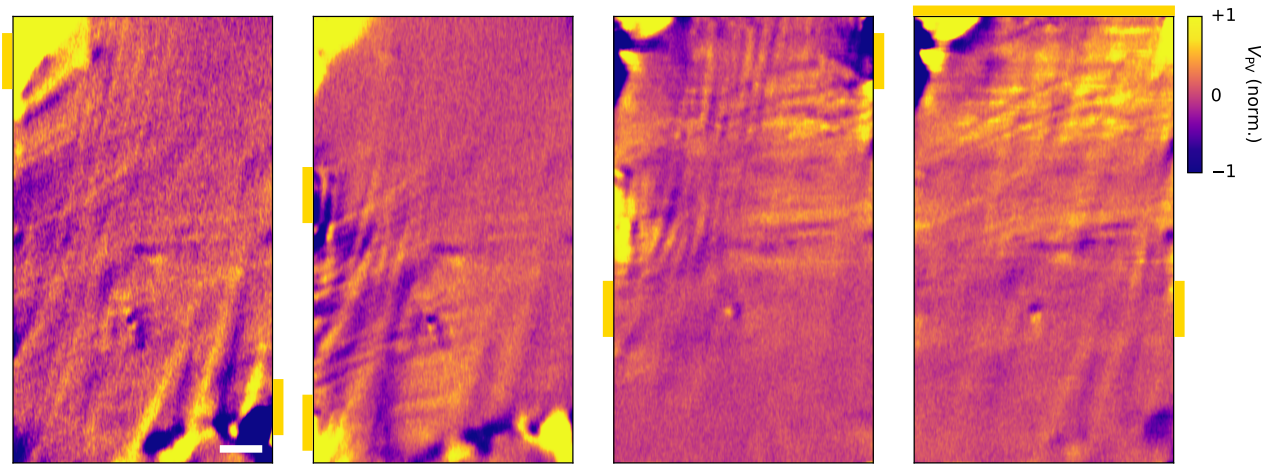}
    \caption{\textbf{Measured photovoltage response with different contact pairs.} Near-field photovoltage maps showing the response for different voltage probes in the undoped sample. The locations of the probes are highlighted by the yellow boxes for each map. $T=10$~K and the scale bar is 1~$\mu$m.}
    \label{fig:app-different-contacts}  
\end{figure*}

\noindent
The measured photovoltage signal is strongly dependent on the choice of voltage probes as described by the Shockley-Ramo theorem\cite{ma2022photocurrent}. This holds for any local voltage source and predicts the global voltage readout. For example, when the photoresponse is driven by the photothermoelectric effect, this is governed by the vector projection of the local gradient of the Seebeck coefficient onto the current paths \cite{Hesp2021}. As a consequence, a different pair of voltage probes causes other areas in our device to "light up" in our experiment, provided there are variations in the Seebeck coefficient. This effect is shown in Figure~\ref{fig:app-different-contacts}, where we show maps of $V_{\rm PV}$ across different voltage probes. Importantly, we observe the same patterns attributed to the second-order superlattice in each maps. All together, these data show that the second-order superlattice is present across the entire bulk of our sample.

\section{Photovoltage response at different temperatures}
\label{sec:app-temperature-dependence} 
\begin{figure}[!b]
    \centering
    \includegraphics{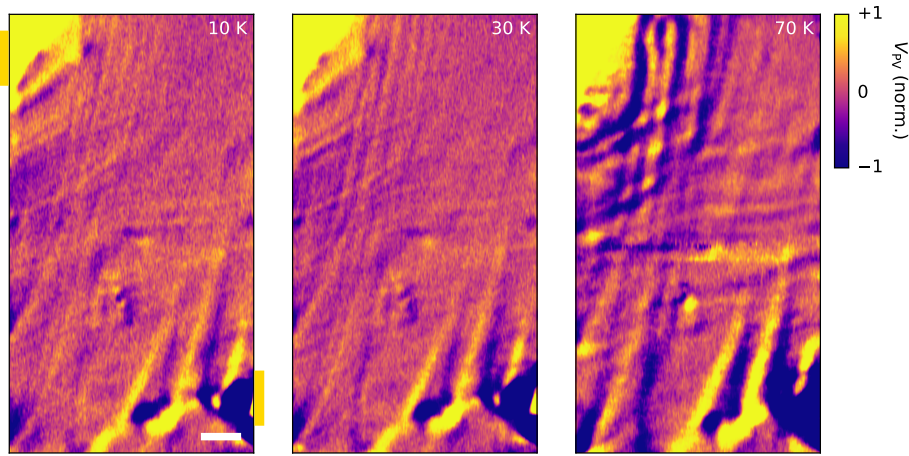}
    \caption{\textbf{Temperature independence of the observed photovoltage features and broken inversion symmetry.} Photovoltage evolution as a function of temperature (T = 10, 30 and 70 K) indicates a rigid background potential, which exhibits no dependence on the temperature. The scale bar is 1 $\mu$m. The locations of the voltage probes are highlighted in yellow.}
    \label{fig:app-temperature-dependence}  
\end{figure}
\noindent
In this Note we provide additional evidence for the observed features originating from a modulating background potential. As shown in Fig.~\ref{fig:app-temperature-dependence}, we map the photovoltage response in the bulk of the sample at 10~K, 30~K, and 70~K. Here, we observe that the overall photovoltage layout persists. This implies that the observed photovoltage patterns are originating from a modulating temperature-independent background potential. This is in contrast to the correlated states, that typically rapidly change their properties with temperature \cite{Cao2018_SC, Cao2018_corr}. Furthermore, we validated that the observed features are also independent of the chosen pair of voltage probes (see Supplementary Note~\ref{sec:app-different-contacts}).

\section{Spatial resolution limit in our experiment}
\label{sec:app-calculation-spatial-resolution}
\noindent
The spatial resolution $L_{\rm res}$ of our experiment is limited by a combination of the AFM tip radius $L_{\rm tip}$ and the inherent spreading $L_{\rm opto}$ attributed to the optoelectronic response, whichever is dominating. Any features below $L_{\rm res}$ will go unresolved, as they are effectively blurred out.

\begin{figure}[!h]
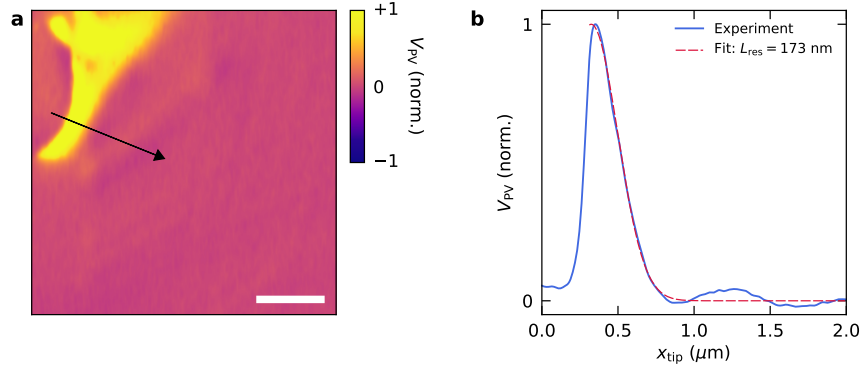

    \centering
    \includegraphics{\pathfigs /SI-spatial-resolution-a.pdf}
    \includegraphics{\pathfigs /SI-spatial-resolution-b.pdf}
    \caption{\textbf{Extraction of spatial resolution of our experiment.} \textbf{a}, Zoom-in of the photovoltage map of Fig.~\NTblue{2}\textbf{a} with an extended color scale. A strong photovoltage is generated at the contact interfacing with MATBG. The scale bar is 1~$\mu$m. \textbf{b}, Line trace of the photovoltage taken along the black arrow in panel \textbf{a}, where the peak position is located slightly away from the contact interface because of the finite tip radius. A fit of the photovoltage profile yields a Gaussian spread of $L_{\rm res} = 173$~nm.}
    \label{fig:app-spatial-resolution}  
\end{figure}

To estimate if $L_{\rm res}$ is independent of the photoresponse mechanism, we use the strong optoelectronic response near the top-left contact as depicted in Figure~\ref{fig:app-spatial-resolution}a. We assume that this local response can be attributed to the sharp metal-graphene interface, and that all the observed spreading of the signal is a result of the finite tip radius and inherent spreading due to the photoresponse mechanism. In Figure~\ref{fig:app-spatial-resolution}b, we fit the data on the graphene-side of the interface with a Gaussian profile and extract $L_{\rm res} = 173$~nm (defined as the standard deviation). We account for the finite spatial resolution in our experiment by filtering the calculated second-order superlattice potentials with a Gaussian profile with a standard deviation of $\sigma = 180$~nm.

\section{Additional transport data}
\label{sec:app-add-transport-data}
\noindent
\begin{figure}[!b]
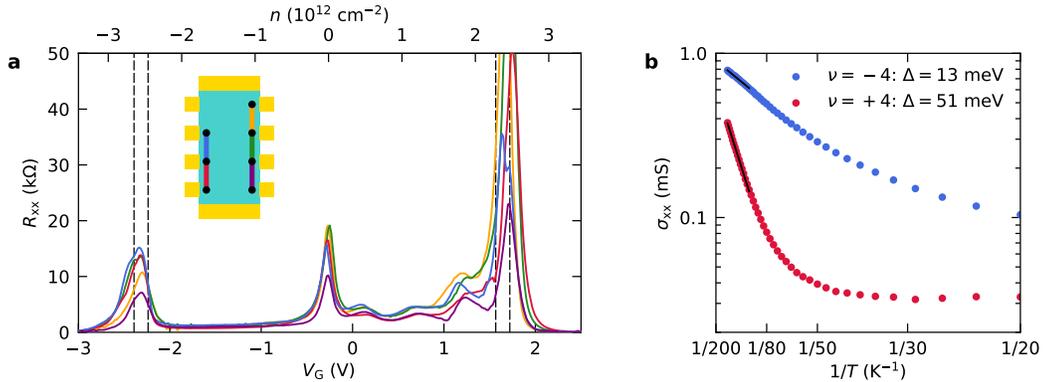

    \centering
    \includegraphics{\pathfigs /SI-transport-a.pdf}
    \includegraphics{\pathfigs /SI-transport-b.pdf}
    \caption{\textbf{Transport characterization of device homogeneity and thermally activated gap.} \textbf{a}, Longitudinal device resistance as measured in a 4-probe configuration. Here we source the current between the bottom-left and top contacts and probe between five different pairs as indicated with the colors (except for the red trace, where the current flows between the bottom-right and top contact). The dashed lines corresponding to twist angles between $1.03 \pm 0.02 \degree$. Data have been obtained at 5~K. \textbf{b}, \NT{Arrhenius plot of the longitudinal conductivity at the full-filling states, taken from the dataset presented in Fig.~\NTblue{3}a. A fit reveals a much smaller gap at $\nu=-4$ compared to $\nu=+4$.}}
    \label{fig:app-transport}  
\end{figure}

Here we discuss additional electronic transport data in our sample. Fig.~\ref{fig:app-transport}a shows four terminal resistance taken for different voltmeter probes available in the studied sample. In this dataset we can identify the same resistance peaks as in Fig.~\NTblue{3}a, with the only difference that the charge neutrality point is off-centered by 90~mV instead of 60~mV. Using this data, we find a mesoscopic twist angle variation of $\sim 1.03 \pm 0.02 \degree$.

Fig.~\ref{fig:app-transport}b depicts the Arrhenius plot for the resistance peak at $\nu \pm 4$. \NT{Thermal activation of the band gap allows us to determine the gap size $\Delta$ via $\sigma_{\rm xx} \propto e^{\Delta/2k_{\rm B}T}$. We find a strongly reduced gap size on the hole side. We note that a more complex fit that would include the nearby resistance peak due to hBN-graphene alignment would lead to an even lower estimate of the gap size at $\nu = \pm 4$. This is expected for our homogeneous device since the resistance can be treated as a sum of the individual resistance peaks (of the full-filling states from the graphene-graphene and hBN-graphene lattices) instead of treating them as originating from parallel channels flowing through our device.}

\NT{
\section{Comparison with the non-aligned MATBG sample}
\label{sec:app-c11-device}
\noindent
To confirm the results in the main text, we performed additional measurements on a different MATBG device, where the two hBN layers are not in close alignment with any of the two graphene layers. From optical microscopy and AFM inspection, we find twist angles of $\sim7 \degree$ and $\sim10 \degree$ between the MATBG and hBN. Electronic transport measurements (Fig.~\ref{fig:app-c11-device}a) show that the device has a twist angle of $1.01 \degree$, in line with the presence of resistance maxima at integer filling factors.

From the electronic transport measurements and cryogenic photovoltage measurements (Fig.~\ref{fig:app-c11-device}), we observe that all the signatures related to the second-order superlattice are not detectable in this device. Namely, the insulating state at charge-neutrality and additional insulating peaks near $\nu=\pm4$ are not present, both of which are otherwise linked to the hBN-graphene superlattice. Likewise, the resistance peaks at both electron and hole integer filling factors indicate that the system exhibits a much stronger electron-hole flat band symmetry. Furthermore, we find that the band gaps at $\nu=\pm 4$ have similar energy gaps (28 and 36 meV for the hole and electron sides, respectively).

When we turn to the cryogenic nanoscale photovoltage measurements for different contact pairs and filling factors (Fig.~\ref{fig:app-c11-device}b), we encounter rather featureless photovoltage maps for a carrier density near the CNP. This confirms that the close alignment between hBN with MATBG is critical for the formation of a second-order superlattice. Even when inspecting the device at an intermediate filling factor (Fig.~\ref{fig:app-c11-device}c), we do not observe any spatially periodic features that we could possibly link to a SOSL. In addition, this does not occur at any other filling factors between $\nu=\pm5$. Nonetheless, we do observe certain modulations in the photovoltage signal that change with the filling factor. The latter indicates that these modulations do not originate from the underlying lattice structure, but rather they relate to the local band structure modulation situated by e.g. local twist angle variations \cite{hesp2021exploring}.

\begin{figure}[!h]
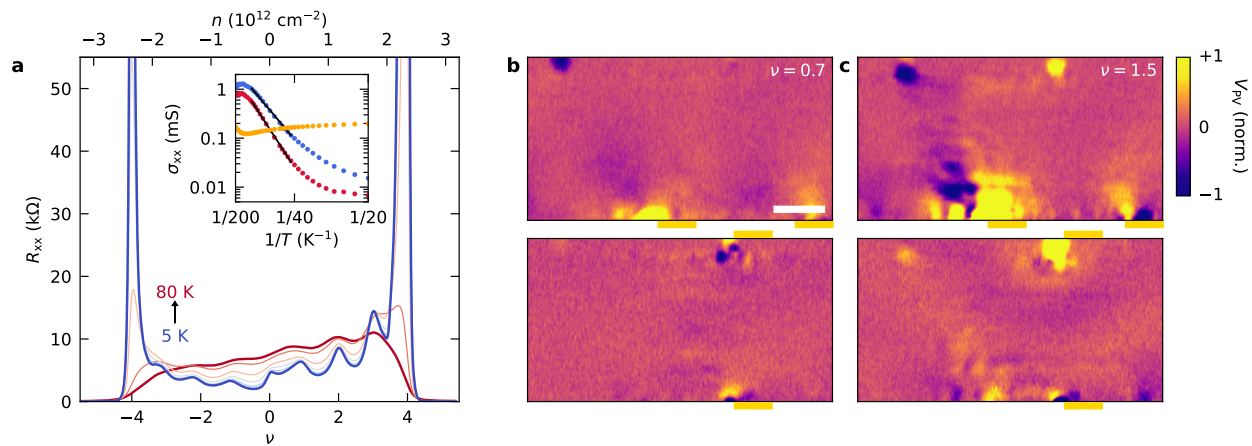

    \centering
    \includegraphics[scale=0.97]{\pathfigs /SI-c11-transport.pdf}
    \includegraphics[scale=0.97]{\pathfigs /SI-c11-PV-maps.pdf}
    \caption{\NT{\textbf{Electronic transport measurements and cryogenic nanoscale photovoltage measurements on a second device.} \textbf{a},~Longitudinal resistivity measurements as a function of MATBG filling factor at different temperatures ranging from 5 to 80 K. Inset: Arrhenius plot of the longitudinal conductivity at the fully-filled or -emptied states (blue: $\nu=-4$, red: $\nu=+4$) and charge-neutrality (yellow). At the CNP, we do not find any signatures of the insulating state, while Arrhenius fit reveals a similar energy gap of 28 and 36~meV at $\nu=-4$ and $\nu=+4$, respectively. \textbf{b}, Near-field photovoltage maps of the entire device taken at $T=10$ K for two different contact pairs at a filling factor $\nu=0.7$. \textbf{c}, Same as panel \textbf{b} but with a higher filling factor. Excitation energy is 115~meV and the scale bar is 2~$\mu$m. Given the weak photovoltage response, we show the first harmonic of the near-field photovoltage signals.}}
    \label{fig:app-c11-device}  
\end{figure}
}

\section{Calculation of the second-order periodicity under strain}
\label{sec:app-periodicity-calculation}
\noindent
To determine the lattice periodicity of the second-order superlattice, we follow the method outlined in the main text. Briefly, we start by calculating the reciprocal vectors of the first-order superlattice as the difference of the original graphene or hBN reciprocal lattice vectors. In turn, the difference of these reciprocal vectors provides us the reciprocal lattice vectors of the second-order superlattice (Fig.~\NTblue{4}a of the main text) \cite{Wang2019}. In the absence of strain, the sixfold rotation symmetry is preserved and therefore these reciprocal lattice vectors have equal length. By taking the inverse of its modulus, and incorporating a factor $2\pi$, we find the second-order lattice periodicity $\tilde{\lambda} _{\rm M}$ (Fig.~\NTblue{4}c of the main text).

\begin{figure}[!h]
    \centering
    \includegraphics{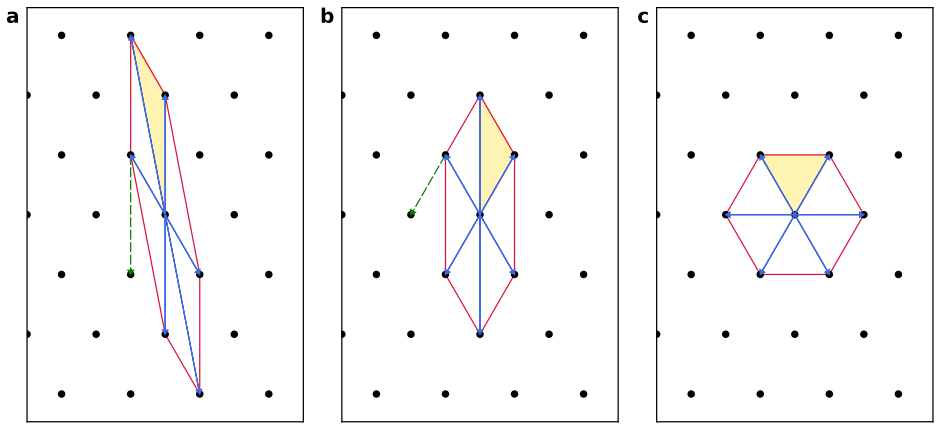}
    \caption{\textbf{Method to find the lattice vectors of the principal unit cell.} \textbf{a}, Example of a triangular lattice indicated by the black dots, together with a set of vectors that do not represent the smallest unit cell. Any unit cell can be composed of a set of triangles (indicated in yellow), which are only acute if they represent the principal unit cell (panel \textbf{c}). To numerically find the lattice vectors that correspond to the principal unit cell, we repeatedly subtract the green dashed vector from the smallest lattice vector to replace the longest vector.}
    \label{fig:app-mapping-triangle}  
\end{figure}

Under the application of strain the second-order reciprocal vectors differ in length (Fig.~\NTblue{4}b of the main text) and we therefore need to alter our approach to determine $\tilde{\lambda} _{\rm M}$. In this case, we need to pay attention to choosing the relevant lattice vector in real space that represents $\tilde{\lambda} _{\rm M}$ (Fig.~\NTblue{4}d in the main text). We start with the set of reciprocal second-order superlattice vectors $\vec{k}_j$ as determined by the method above. As we are to evaluate the real-space periodicity, we calculate the two real-space lattice vectors $\vec{r}_i$ from $\vec{k}_j$ using the relation $\vec{r}_i \vec{k}_j = 2\pi \delta _{ij}$. This yields a complete set of six real-space vectors, as illustrated in Fig.~\ref{fig:app-mapping-triangle}a. A common outcome of our calculation is a set of lattice vectors that does not represent the principal unit cell. The maximum superlattice periodicity that follows from this will be an overestimation with respect to those from the more obvious choice of unit cell (Fig.~\ref{fig:app-mapping-triangle}c).

To correct for this, we define the principal unit cell as the set of vectors that comprise an acute triangle (all angles smaller than $90\degree$). This condition allows us to convert any set of lattice vectors (Fig.~\ref{fig:app-mapping-triangle}a) back to the principal unit cell (Fig.~\ref{fig:app-mapping-triangle}c) via multiple intermediate steps (Fig.~\ref{fig:app-mapping-triangle}b). In this approach, the longest lattice vector (and its opposite) is replaced by the vector defined by the smallest vector subtracted from either the middle or longest vector, whichever yields the smallest vector. This is indicated by the green dashed vector in Fig.~\ref{fig:app-mapping-triangle}a, and yields a reduced unit cell. We repeat this until the set of vectors meet the condition of an acute triangle: $a^2+b^2 < c^2$, where $a$, $b$ and $c$ are the smallest, intermediate and longest vectors. With this new set of real-space lattice vectors, we define the maximum second-order superlattice periodicity $\tilde{\lambda} _{\rm M}$ as the modulus of the longest vector.

\begin{figure}[!h]
    \centering
    \includegraphics{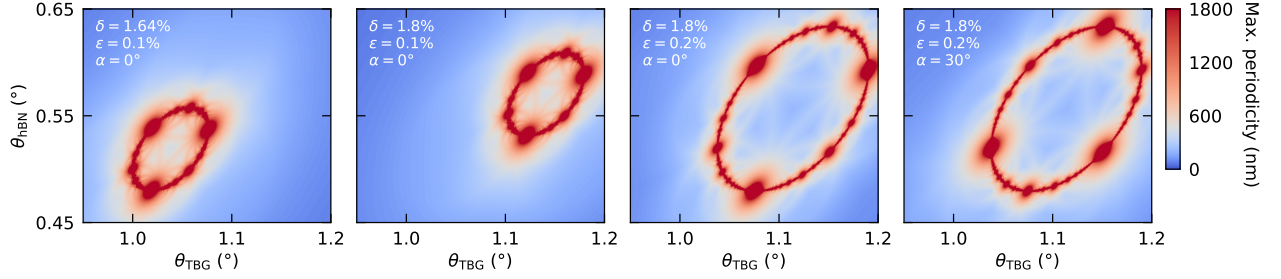}
    \caption{\NT{\textbf{Calculated SOSL periodicity for various model parameters.} Maximum SOSL periodicity calculated for the hBN-TBG system with identical strain applied to both graphene layers. The four panels highlight the influence of a different hBN-graphene lattice mismatch, a different strain magnitude and a different axis along which the strain is applied.}}
    \label{fig:app-periodicity-calculation-mismatch-strain}  
\end{figure}

\NT{
\section{Effect of hBN-mismatch, strain magnitude and angle on SOSL periodicity}
\label{sec:app-periodicity-calculation-mismatch-strain}
\noindent
Our phenomenological model as introduced in Fig.~\NTblue{4} of the main text has several parameters that influence the calculated SOSL periodicity. These parameters include graphene-hBN lattice mismatch $\delta$, strain magnitude $\epsilon$ and strain angle $\alpha$ with respect to the zigzag direction. To explore the effect and sensitivity on these parameters, we repeat the calculation of Fig.~\NTblue{4}d in the main text while tuning each of the mentioned knobs. Fig.~\ref{fig:app-periodicity-calculation-mismatch-strain} shows the results of these calculations, and the second panel corresponds to the Fig.~\NTblue{4}d shown in the main text. We find that adjusting the hBN-graphene lattice mismatch effectively moves the resonance along the line of $\theta_{\rm TBG} = \theta_{\rm hBN}$, until eventually the case of trilayer graphene is approached as in Fig.~\ref{fig:app-trilayer}b at $\delta=0 \%$. Furthermore, increasing the strain magnitude (of the two graphene layers together) increases the extent of the SOSL, meaning that for a wider range of twist angles a SOSL can be observed. Finally, changing the direction along which the strain is applied effectively rotates the fractal-like features around its axis in $\theta_{\rm TBG}$-$\theta_{\rm hBN}$ space. Together with the experimental data, this shows there can be complimentary parameter sets that could account for our experimental observations.}

\section{Calculations of the real-space potentials and requirement for a non-linear response}
\label{sec:app-calculation-real-space-potential}
\noindent
Here, we develop a toy model to visualize the second-order superlattice in real space, which helps us understand its properties as a function of the underlying lattice parameters (see Fig.~\NTblue{5} of the main text). Our model reflects a non-linear interaction between the individual first-order lattice potentials, which we define and describe below.

\begin{figure*}[!b]
    \centering
    \includegraphics[scale=\scalefig]{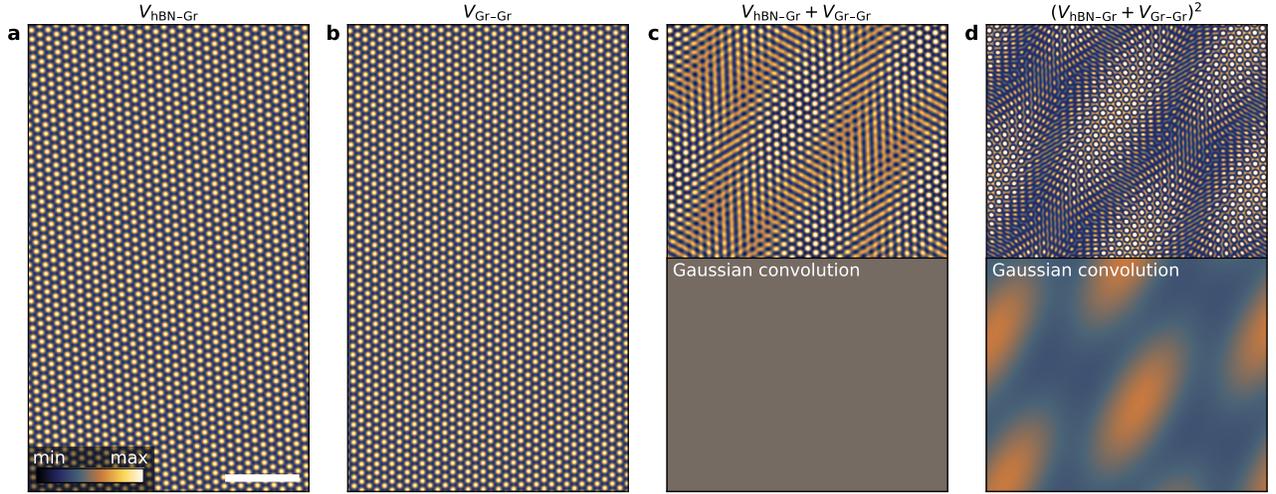}
    \caption{\textbf{Visualization of the second-order superlattice potentials in real space.} \textbf{a}, Calculated real-space potential of the hBN-Gr superlattice. \textbf{b}, Same as \textbf{a} but for the Gr-Gr superlattice. \textbf{c}, Second-order superlattice potential $V_{\rm tot}$ calculated as the sum of the first-order superlattice potentials (panels \textbf{a} and \textbf{b}). The lower panel reflects the combined potential after a Gaussian convolution with a standard deviation of 20~nm. \textbf{d}, Same as \textbf{c} but with $V_{\rm tot}$ defined as the square of the first-order superlattice potentials. As shown in the bottom panel, a Gaussian convolution only filters the short-range modulations away, but preserves the second-order superlattice features. The scale bar is 100~nm. The bounds of the color scales are adjusted in each panel to optimize clarity.}
    \label{fig:app-calculation-potential-superlattice}  
\end{figure*}

First, we calculate the lattice potentials $V_{\rm hBN-Gr}$ and $V_{\rm Gr-Gr}$ of the two first-order superlattices formed by the hBN-Gr and Gr-Gr layers, respectively. We define these potentials to be the interlayer binding energy as a function of the local displacement vector $\vec{\delta} (\vec{r})$ for a given position $\vec{r}$. In doing so, we follow Ref.~\citenum{Nam2017}, which gives $V(\vec{r}) = \sum _{j=1} ^{3} 2V_0 \cos(\vec{k}_j \cdot \vec{\delta}(\vec{r}))$, and set $V_0=1$ for simplicity. $\vec{k}_{1,2}$ are the reciprocal lattice vectors of one of the two layers involved in each superlattice and $\vec{k}_3 = -\vec{k}_1 -\vec{k}_2$). The effect of strain is incorporated following Eq.~(\NTblue{1}) of the main text. Then we decompose the position vector $\vec{r}$ into the real-space superlattice vectors (determined from the reciprocal counterparts, shown as the red vectors in Fig.~\NTblue{4} of the main text) and take their residuals. The ratio of the residuals with respect to the real-space superlattice vectors naturally provides us with the local displacement vector $\vec{\delta} (\vec{r})$ as a fraction of the real-space lattice vectors of one of the two layers involved in each superlattice (the counterparts of $\vec{k}_j$). Figure~\ref{fig:app-calculation-potential-superlattice}a,b show the calculated potentials of the hBN-Gr and Gr-Gr superlattices. Here, we use $\theta_{\rm hBN}=0.48 \degree$ and $\theta_{\rm TBG}=1.08 \degree$, while we keep all the other parameters the same as in the main text.

Next, we initially define the second-order superlattice potential $V_{\rm tot} = V_{\rm hBN-Gr} + V_{\rm Gr-Gr}$ as the sum of the underlying superlattice potentials. As can be seen in Figure~\ref{fig:app-calculation-potential-superlattice}c, it is possible to identify the long-range potential modulations. Since our measurement scheme is only sensitive at length scales beyond the AFM tip radius ($\approx 50$~nm) and the cooling length ($\approx 140$~nm, see Supplementary Note~\ref{sec:app-calculation-spatial-resolution}), our measurements will not be able to resolve any of the features of the first-order superlattice. However, with $V_{\rm tot}$ defined as a mere sum of the underlying potentials, the long-range modulations associated to the second-order superlattice also remain undetectable in our measurements. This is illustrated in the bottom panel of Fig.~\ref{fig:app-calculation-potential-superlattice}c, where $V_{\rm tot}$ is convoluted with a Gaussian filter with a standard deviation $\sigma =20$~nm. We choose this relatively small filter width for illustrative purposes, while Fig.~\NTblue{5} depicts the case close to the experimental parameters for $\sigma =140$~nm. The fact that we observe a flat profile in Fig.~\ref{fig:app-calculation-potential-superlattice}c is not surprising itself, since the Gaussian filter acts directly on the first-order potentials, which have a superlattice periodicity smaller than $\sigma$.

To fit our observations within the model of a second-order superlattice, it is required to preserve long-range potential modulations upon spatial smearing. The simplest and most reasonable approach towards this is by coupling the first-order superlattice potentials in a non-linear fashion. To illustrate this, we repeat the same calculations, but with $V_{\rm tot} = (V_{\rm hBN-Gr} + V_{\rm Gr-Gr})^2$ (Fig.~\ref{fig:app-calculation-potential-superlattice}d). Such coupling introduces terms that are insensitive to the Gaussian convolution, as can be seen in the bottom panel of Fig.~\ref{fig:app-calculation-potential-superlattice}c. 
In more general terms, we argue that the optoelectronic response should inherently possess one or more non-linearities in its pathway and does not necessarily have to occur in the coupling between the two superlattices. While the actual interaction between the two lattice potentials is beyond the scope of this work, we can speculate on a multitude of effects that could introduce such non-linearity: 1) an interaction of the two first-order potentials that is not linear (as discussed above), possibly due to the inequality of hBN and graphene or due to the lattice reconstruction; 2) the dependence of the Seebeck coefficient on the local electronic properties (including the potential landscape). These examples concern the fundamental question about how the two superlattices interact and define the properties of the resulting second-order superlattice.

\section{Calculations of SOSL in twisted trilayer graphene} 
\label{sec:app-trilayer}
\begin{figure*}[b!]
    \centering
    \includegraphics[scale=\scalefig]{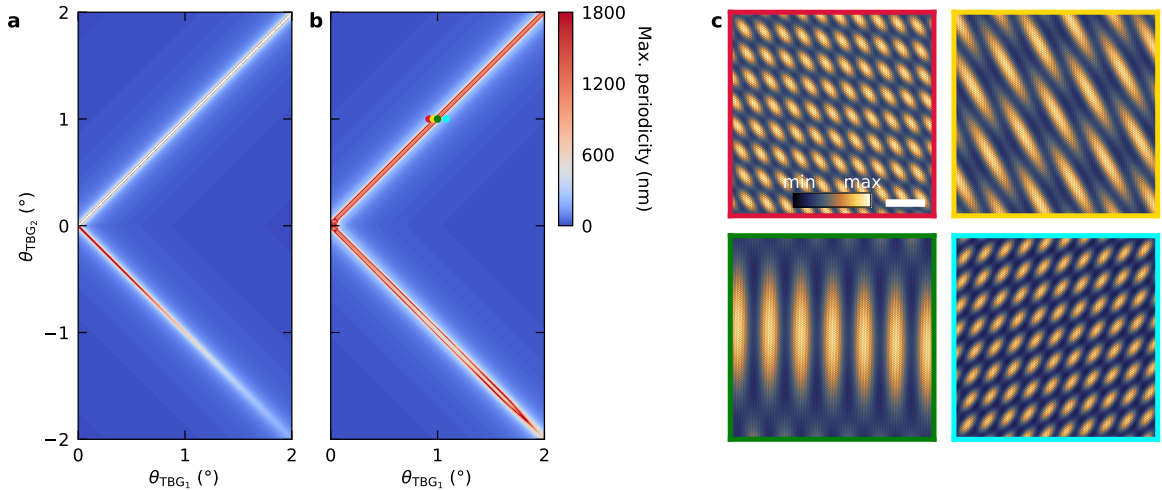}
    \caption{\NT{\textbf{Second-order superlattices in twisted trilayer graphene.} \textbf{a}, SOSL periodicity as a function of $\theta_{\rm TBG_1}$ and $\theta_{\rm TBG_2}$ of the unstrained SOSL lattice. \textbf{b}, Maximum SOSL periodicity (along one of its principal axes) of the deformed lattice when strain applied to the upper two graphene layers. \textbf{c}, Four examples of different SOSL geometries, including a square lattice and a quasi-1D lattice. The color of the frames corresponds to the color of the dots in \textbf{b}. The color bar represents the calculated potential, and we note that the underlying first-order superlattices are visible. For clarity we smoothed the data with a Gaussian filter of width $\sigma=3.5$~nm ($\sigma$ is standard deviation of Gaussian kernel). The scale bar is 250~nm.}}
    \label{fig:app-trilayer}  
\end{figure*}

\noindent
\NT{The second-order superlattice reported in the main text is not limited to a system of twisted bilayer graphene in close alignment with hBN. In principle, any pair of superlattices forms a second-order superlattice, provided that the underlying superlattices have similar periodicities. To extend the discussion about the effect reported in the main text, here we present additional theoretical modelling for twisted trilayer graphene (see Fig.~\ref{fig:app-trilayer}). Just as we observed in the hBN-MATBG system, also in trilayer graphene there are certain combinations of $\theta_{\rm TBG_1}$ and $\theta_{\rm TBG_2}$ (the twist angles between the bottom/middle and middle/top graphene layers) where long-range lattice potential modulations are expected. Yet, the range of twist angles for which this occurs is not as limited as in the case of hBN-TBG, but appears around the condition $\theta_{\rm TBG_1} = \pm \theta_{\rm TBG_2}$. This can be understood from the fact that two pairs of graphene lattices at the same or opposite twist angle have the first-order superlattice in common. This directly satisfies the condition of the two underlying superlattices being aligned with each other in order to create a second-order superlattice. It is worth noting that this coincides with the observation of exotic states in twisted trilayer graphene, occurring at $\theta_{\rm TBG_1} = -\theta_{\rm TBG_2} \approx 1.57 \degree$ \cite{park2022robust}.}

\section{Simulation of spatial photovoltage response}
\label{sec:app-simulation}
\begin{figure*}[b!]
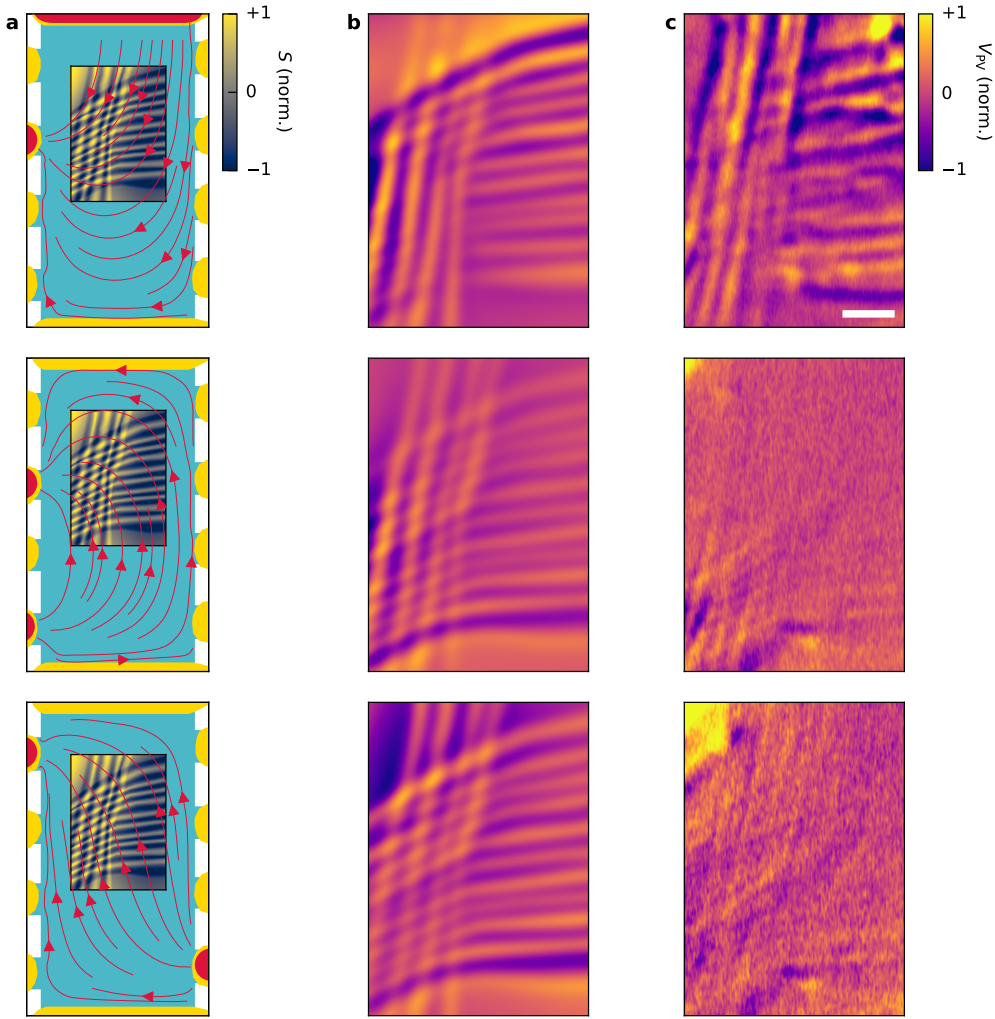

    \centering
    \includegraphics[scale=\scalefig]{\pathfigs /Fig-SI-simulation-current-flows-1.pdf}
    \includegraphics[scale=\scalefig]{\pathfigs /Fig-SI-simulation-response-1.pdf}
    \includegraphics[scale=\scalefig]{\pathfigs /Fig-SI-simulation-experiment-1.pdf}
    \includegraphics[scale=\scalefig]{\pathfigs /Fig-SI-simulation-current-flows-2.pdf}
    \includegraphics[scale=\scalefig]{\pathfigs /Fig-SI-simulation-response-2.pdf}
    \includegraphics[scale=\scalefig]{\pathfigs /Fig-SI-simulation-experiment-2.pdf}
    \includegraphics[scale=\scalefig]{\pathfigs /Fig-SI-simulation-current-flows-3.pdf}
    \includegraphics[scale=\scalefig]{\pathfigs /Fig-SI-simulation-response-3.pdf}
    \includegraphics[scale=\scalefig]{\pathfigs /Fig-SI-simulation-experiment-3.pdf}
    \caption{\textbf{Simulations of the photo-thermoelectric response.} \textbf{a}, Geometry of our device with the current lines indicated by the red arrows and running between three different pairs of contacts (marked by the red color). The inset marks the region of interest and shows the Seebeck coefficient. This serves as input parameter of our simulations. \textbf{b}, Calculated photovoltage within the region of interest corresponding to the three configurations of the photovotage probes in panel \textbf{a}. Here, we follow the simulation method as detailed in Ref.~\citenum{Hesp2021}. \textbf{c}, Measured photovoltage response for the three corresponding orientations of the contacts. The experimental data are partially cropped from those shown in Fig.~\NTblue{1}b of the main text, and the two left panels of Fig.~\ref{fig:app-different-contacts}. The scale bar is 1~$\mu$m.}
    \label{fig:app-simulations}  
\end{figure*}
\noindent
To gain further insights into the measured photovoltage response and its relation to the underlying second-order superlattice structure, we perform simulations of the expected photovoltage following the same method as used in Ref.~\citenum{Hesp2021}. Our model simulates the photothermoelectric response and provides the photovoltage responsivity for a given spatial profile of the Seebeck coefficient and electronic conductivity.

Figure~\ref{fig:app-simulations}a shows the model parameters for the three cases that differ only by the pair of the measurement contacts, leading to a marked difference of the current flow within the device. To define the spatial profile of the Seebeck coefficient $S$, we apply the following strategy: first, guided by the measured photovoltage maps, we construct two sets of lines (semi-horizontal and semi-vertical), which together govern the second-order superlattice structure. Second, for each set of lines, we define a cosine function crossing zero at each of the lines, which smoothly approaches zero towards the sample edges. Finally, the sum of these two cosine functions forms a continuous representation of $S$ serving as an input of the simulations (see insets in Fig.~\ref{fig:app-simulations}a). All other input terms of the simulation are constant: the temperature is fixed at $10$~K, the electrical conductivity is $1$~mS, the cooling length $L_{\rm c}$ is set to $140$~nm. After specifying these input values, we now calculate the photovoltage responsivity. The simulated photovoltage profile follows by convolving the responsivity with a Gaussian spread function that accounts for the finite tip size injecting heat into MATBG. We set the Gaussian width $\sigma$ to $100$~nm, such that $\sqrt{L_{\rm c}^2 + \sigma ^2} = 173$~nm, equalling the spatial resolution in our experiment as determined in the Supplementary Note~\ref{fig:app-spatial-resolution}.

The simulated photovoltage is shown in Fig.~\ref{fig:app-simulations}b. At the first glance, we find that the calculated photovoltage response mimics the Seebeck profile. Yet, at a closer inspection, we observe that the three simulations each highlight different parts within the region of interest showcasing locally a stronger contrast and higher magnitude of the fringes. In addition, depending on the local orientation of the current flow, we find that either the semi-horizontal or semi-vertical fringes appear more pronounced for the three different cases. This is explained by noting that the photovoltage responsivity can be described as a projection of the gradient of the inhomogeneous Seebeck coefficient onto the local current flow \cite{ma2022photocurrent, Song2014}. Finally, we observe that the magnitude of the photovoltage depends on the distance between the fringes, which can be explained by the finite cooling length and Gaussian spread function, which both reduce the magnitude of fringes that are close to one another.

When comparing these simulations with the corresponding experimental data in Fig.~\ref{fig:app-simulations}c, we find a good qualitative agreement. In particular, the above mentioned characteristics following from the different choices of the contact probes are clearly reflected. This strongly indicates that the chosen input of $S$ shows the actual SOSL structure, which we classify in the main text as a square lattice (see Fig.~\NTblue{5}b of the main text). On the other hand, we also observe that the measured signal in the middle row decays faster towards the top-right corner than in the simulation. We hypothesize that a more complex spatial conductivity profile could account for this difference. Given the large resistivity variations in MATBG as function of doping and local twist angle such scenario is highly probable.

\section{Discussion on the photovoltage generation mechanism}
\label{sec:app-PV-mechanism}
\begin{figure*}[t!]
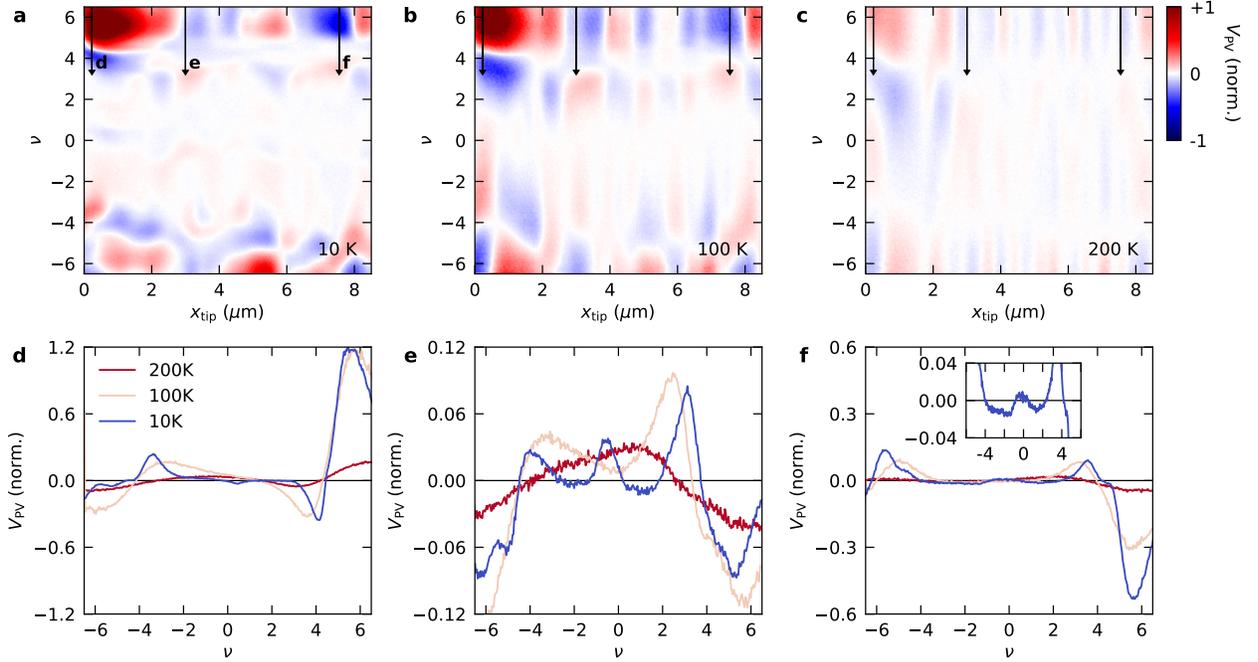

    \centering
    \includegraphics[scale=\scalefig]{\pathfigs /Fig-SI-gatetrace-temperature-abc.pdf}
    \includegraphics[scale=\scalefig]{\pathfigs /Fig-SI-gatetrace-temperature-def-traces.pdf}
    \caption{\NT{\textbf{Dependence of photovoltage signal on temperature $T$ and filling factor $\nu$.} \textbf{a}-\textbf{c}, Measurements of $V_{\rm PV}$ along the vertical trace across the sample reported in the main text for a range of filling factors. This is repeated at three different temperatures (10~K, 100~K, 200~K). \textbf{d}-\textbf{f}, Line traces of $V_{\rm PV}$ along the indicated arrows indicated in panels \textbf{a}-\textbf{c}. Panel \textbf{d} corresponds to a position nearby the top contact of the device, while panel \textbf{e} and \textbf{f} are in the bulk of the sample. All the presented data are normalized to the same value. The inset in panel \textbf{f} is a vertical zoom of the trace at 10~K.}}
    \label{fig:app-temperature-gate-dependence}  
\end{figure*}

\noindent
In many cases, the optoelectronic response in graphene-based devices is primarily facilitated by the photothermoelectric effect (PTE), which drives a photovoltage that is proportional to the spatial gradients in the Seebeck coefficient $S$ \cite{Gabor2011, Lemme2011, Woessner2016, Hesp2021, ma2022photocurrent}. In our case, it may play a role in the photoresponse pathway, since the Seebeck coefficient $S$ is a function of the electrical conductivity, which in turn is governed by the electronic properties of the atomic lattice. Hence, this pathway provides a projection of the underlying lattice structure on the photovoltage maps. In this scenario, we can explain the strong enhancement of the photovoltage response for higher filling factors (as seen in Fig.~\NTblue{2}d) by an increased cooling length due to the enhanced charge carrier mobility in the remote bands \NTrev{\cite{Dudley2024}}, as well as by the enhanced absorption efficiency \cite{Deng2020}. In the vicinity of more mobile charge carriers (i. e. outside the flat band regime), the photovoltage electrodes become less sensitive to the spatial variations of the Seebeck coefficient, while the reduced charge carrier mobility inside the flat band boosts our probes' spatial sensitivity.

Figure~\ref{fig:app-temperature-gate-dependence} provides an in-depth view of the dependence of the photovoltage response for a wide range of the filling factors and temperatures between 10~K and 200~K. Fig.~\ref{fig:app-temperature-gate-dependence}d-f show line traces for a direct comparison between different temperatures. For the traces obtained near the contact (Fig.~\ref{fig:app-temperature-gate-dependence}d), the photovoltage response due to the PTE would be approximately proportional to the local Seebeck coefficient in the TBG system \cite{hesp2021exploring}. Partially in line with this, at 10~K we retrieve a typical profile given by the Mott formula, as seen by the sign changes near the resistance peak (at $\nu\pm4$ and charge neutrality). Such profile is also shown in Fig.~\ref{fig:app-temperature-gate-dependence}f. However, in the bulk of the sample, represented by rather different profiles of the traces in Fig.~\ref{fig:app-temperature-gate-dependence}e, it is harder to establish whether the PTE is dominating the photovoltage response. Very small twist angle variations may also be responsible for producing such profiles in conjunction with the PTE \cite{hesp2021exploring}. The observed weak reduction (or even enhancement) of the signal between 10~K and 100~K is something possible within the PTE, but not expected (as for instance in good metals), as it includes many multiple temperature dependencies (for instance, the cooling length and the Seebeck coefficient).

Another possible photovoltage generation mechanism is the second-order photoresponse, which has been reported to be more significant in the flat band systems \cite{chaudhary2022shift, ma2022intelligent, arora2021strain, kaplan2022twisted}. This type of photoresponse is driven by a momentum difference between the initial and excited states, or a real-space wave function shift upon an excitation \cite{ma2022photocurrent}. Due to its nature, the second-order photoresponse depends strongly on the quantum geometry (such as the Berry curvature of the band structure) of the system under study. In particular, a second-order photoresponse can be enabled when the inversion symmetry is broken due to close alignment of TBG with the hBN lattice. On the other hand, uniaxial strains can lead to a non-zero bulk photoresponse for unpolarized light. Specifically, infrared interband transitions in twisted bilayer graphene have been found to produce a dominant second-order photocurrent response (linear shift current)\cite{ma2022intelligent, ma2022photocurrent}. 

Unfortunately, a comprehensive understanding of the second-order photovoltage generation mechanism is not trivial because its magnitude and dependencies on charge carrier density are hidden in the geometric properties of the bands. As demonstrated in this work, even at these high energy scales (the excitation energy of 116 meV with respect to the interband transitions) the photoresponse can vary quite dramatically with doping. Regarding the temperature dependence, the second-order bulk photovoltaic response depends on the relative population between bands and hence becomes quickly suppressed when the temperature approaches the equivalent energy scales of interband transitions. Indeed, previous works have demonstrated a monotonic decrease in the second-order response with temperature\cite{ma2022intelligent} that contrasts with our measured non-monotonic dependence potentially pointing towards a PTE dominated mechanism. However, both $T$- and charge carrier density dependencies share many similarities and the only way to truly distinguish between the two effects is via polarization dependent measurements, for which the PTE is insensitive whilst the second-order response is highly sensitive. These of course are not possible in the near-field measurements and hence we leave the study of the second-order photocurrents in these types of heterostructures for future works.

\newpage
\noindent
\bibliographystyle{naturemag-achim}
\bibliography{supplement.bib}